\documentclass[twocolumn]{aastex7}
\usepackage{caption}
\usepackage[skip=0.5ex]{subcaption}
\usepackage{amsmath,amstext}
\usepackage[T1]{fontenc}
\usepackage{apjfonts}
\usepackage[figure,figure*]{hypcap}
\usepackage{multirow}
\usepackage{subfiles}
\usepackage{graphicx}
\usepackage{xspace}
\DeclareGraphicsExtensions{.pdf,.png}

\newcommand{\Aeos}{A{\sc eos}\xspace}
\newcommand{\RomanNumeralCaps}[1]{\MakeUppercase{\romannumeral #1}}

\defcitealias{Emerick2019}{E19}

\begin{document}

\title{From Primordial Stars to Early Galaxies: A Semi-Analytic Model Calibrated with \textsc{Aeos} and Renaissance}

\correspondingauthor{Ryan Hazlett}
\email{ryan.hazlett@rockets.utoledo.edu}


\author[0000-0002-1034-7986]{Ryan Hazlett}
\affiliation{Ritter Astrophysical Research Center, Department of Physics and Astronomy, University of Toledo, 2801 W. Bancroft Street, Toledo,
OH 43606, USA}
\email{ryan.hazlett@rockets.utoledo.edu}

\author[0009-0006-4744-2350]{Jennifer Mead}
\affiliation{Department of Astronomy, Columbia University, New York, NY 10027, USA}
\email{jennifer.mead@columbia.edu}

\author[0000-0002-8365-0337]{Eli Visbal}
\affiliation{Ritter Astrophysical Research Center, Department of Physics and Astronomy, University of Toledo, 2801 W. Bancroft Street, Toledo,
OH 43606, USA}
\email{elijah.visbal@utoledo.edu}

\author[0000-0003-2630-9228]{Greg L. Bryan}
\affiliation{Department of Astronomy, Columbia University, New York, NY 10027, USA}
\affiliation{Center for Computational Astrophysics, Flatiron Institute, 162 5th Ave, New York, NY 10010, USA}
\email{greg.bryan@columbia.edu}

\author[0000-0003-0064-4060]{Mordecai-Mark Mac Low}
\affiliation{Department of Astrophysics, American Museum of Natural History, New York, NY 10024, USA}
\affiliation{Department of Astronomy, Columbia University, New York, NY 10027, USA}
\email{mordecai@amnh.org}

\author[0000-0002-9789-6653]{Mihir Kulkarni}
\affiliation{Institut f{\"u}r Astrophysik und Geophysik, Georg-August Universit{\"a}t G{\"o}ttingen, Friedrich-Hund-Platz 1, 37077 G{\"o}ttingen, Germany}
\email{mihir.kulkarni@uni-goettingen.de}

\author[0000-0003-3479-4606]{Eric P. Andersson}
\affiliation{Department of Astrophysics, American Museum of Natural History, New York, NY 10024, USA}
\email{eandersson@amnh.org}

\author[0000-0002-8810-858X]{Kaley Brauer}
\affiliation{Center for Astrophysics | Harvard \& Smithsonian, Cambridge, MA 02138, USA}
\email{kaley.brauer@cfa.harvard.edu}

\author[0000-0003-1173-8847]{John H. Wise}
\affiliation{Center for Relativistic Astrophysics, Georgia Institute of Technology, 837 State Street, Atlanta, GA 30332, USA}
\email{jwise@physics.gatech.edu}


\begin{abstract}
We present an extension of our semi-analytic model that follows the formation of Population III stars and their metal-enriched descendants, incorporating dark matter halo merger trees from cosmological $N$-body simulations and feedback from reionization. Our extended model is calibrated using two complementary cosmological hydrodynamical simulations: \Aeos, which resolves individual Population III and II stars to $z\sim14.6$, and Renaissance, which is lower resolution but follows large-scale metal-enriched star formation to $z \sim 11$. With a combined calibration, we capture small-scale physics of primordial star formation over a large range in halo mass. We find good agreement between our calibrated model and \Aeos, reproducing the evolution in number of star-forming halos and total stellar mass. Achieving this agreement requires increasing the normalization of, flattening the redshift dependence of, and adding scatter to the commonly used critical mass threshold $M_{\mathrm{crit}}$. Our treatment of the delay between Pop III stellar death and subsequent Pop II star formation emphasizes the need to account for halos that have yet to transition to Pop II, since incomplete sampling of this delay in simulations limits physically motivated calibrations. Finally, we apply our model to larger-volume dark matter only simulations and predict $\sim10$ active Pop III sources at $z = 10$ lie within the area strongly lensed by galaxy cluster MACS J0416 with a magnification exceeding $\mu > 30$. These results demonstrate that semi-analytic approaches, when calibrated to hydrodynamical simulations, can provide accurate, computationally efficient predictions for the earliest stages of cosmic star formation.
\end{abstract}

\section{Introduction}
The direct descendants of the first stars, the most metal-poor stars that we observe in our own Milky Way halo and in the Local Group, currently provide the best observational insight into the properties of the first metal-free stars, designated Population (Pop) III, and the first galaxies that formed at redshifts $30<z<20$ \citep{FrebelNorris2015,KlessenGlover2023}. Although they existed only briefly in each galaxy’s evolution, Pop III stars marked the onset of cosmic chemical evolution by forming the first elements heavier than Li. These newly forged metals mixed into the surrounding gas, setting the initial abundance patterns inherited by the next generation of metal-poor Pop II stars.

Direct observations of Pop III stars remain elusive due to their short lifetimes and formation predominately at high redshift. However, recent results suggest that their signatures may now be within reach. Rare cases of strong gravitational lensing can amplify faint stellar clusters to observable levels \citep{Windhorst2018,Schauer2020,Vanzella+2023,Adamo+2024, Mowla+2024}. The lensed system LAP1-B at $z=6.6$ shows multiple hallmarks of Pop III star clusters: formation in pristine halos and a top-heavy initial mass function \citep{Nakajima2025,Visbal2025b}. Several high-redshift galaxies exhibit strong He \RomanNumeralCaps{2} $\lambda 1640$ emission or very low metallicities \citep{Vanzella2023,Maiolino2024,Wang2024,Fujimoto2025,Cai2025,Morishita2025}, consistent with signatures related to Pop III stars.

There is also a range of complementary observables that provide indirect constraints on the first stars. These include transient phenomena that probe Pop III remnants, including supernovae \citep[SNe;][]{Hartwig2018,Moriya2019} or gamma-ray bursts \citep{Toma2016}, and gravitational waves from black hole mergers \citep{Inayoshi2016, Safarzadeh2020, Liu2024}. Absorption studies in objects like the recently identified damped Lyman-$\alpha$ systems (DLAs) at $z \sim 6$ show abundance ratios, particularly  [C/O] and [Si/O], that point to enrichment from Pop III SNe \citep{Sodini2024,Visbal2025a}. Global measurements reflecting Pop III star formation include the optical depth of the cosmic microwave background \citep{Visbal2015b}, the cosmological 21-cm signal \citep{Mirocha2018,Gessey-Jones2022}, and He II $\lambda 1640$ line intensity mapping \citep{Visbal2015a, Parsons2022}. Finally, stellar archaeology can reveal how Pop III SN ejecta are imprinted on the next generation of metal-poor stars \citep{FrebelNorris2015}. Interpreting these diverse observational probes requires a theoretical framework for the properties of the first stars. Of particular interest is constraining the uncertainty in the Pop III initial mass function (IMF), their lifetimes, feedback, and chemical enrichment,  which imprint their signatures on subsequent generations of stars and galaxies.

The Pop III IMF is thought to be top-heavy due to the slow cooling of primordial clouds through the H$_2$ rovibrational line \citep{Haiman1996-2,Tegmark1997}, which would result in large Jeans masses and stellar masses on the order of $10^2$--$10^3\,M_\odot$ \citep{Abel2002,Bromm2002,Bromm2004,Yoshida2008,Hirano2015,Hosokawa2016}.  Other recent theoretical work has demonstrated that Pop III stars may actually form with characteristic masses of tens of solar masses or lower \citep{Stacy2010,Clark2011,Greif2011,Greif2012,Hosokawa2011,Hosokawa2016,Latif2013,Latif2022,Hirano2014,Sharda2020}, or that they may even form with masses below $0.8\,M_\odot$ \citep{Jaura2022,ShardaMenon2024preprint,Lake2024preprint}. An IMF with a power-law slope shallower than \citet{Salpeter1955} is supported by observations of metal-poor dwarf galaxies \citep{Geha2013}, which support the hypothesis of an evolving IMF that was bottom-light at early times. A flat IMF for massive stars (10--300~$M_\odot$) is disfavored by observations of carbon-enhanced metal-poor stars \citep{deBennassuti2017}.  Additionally, a few individual metal-poor stars have been observed with abundances that suggest that they have been enriched by a single SN \citep{Ezzeddine2019,Skuladottir2021,Skuladottir2024,Yong2021,Xing2023,Ji2024}, and could be used to provide constraints on the yields of Pop III stars. However, the interpretation of these abundances and their connection to chemical yields is sensitive to the retention and mixing of ejected metals into the interstellar medium (ISM, see, e.g., \citealt{Mead+2025a, Mead+2025b}).

Understanding the observations of extremely metal-poor stars and their implications for the properties of Pop III stars and galaxies requires computational models of galaxy formation and evolution complete from the first stars to the present day.  One avenue to pursue is through numerical hydrodynamic simulations that follow the evolution of dark matter (DM) and star particles, as well as gas.  In particular, studying the formation, evolution, and effects of Pop III stars requires high spatial and mass resolution in simulations that model individual stars \citep{Emerick2019,Lahen2020,Lahen+2023,Gutcke+2021,Andersson2023,Andersson2025,AeosMethods} in order to capture sites of Pop III star formation in minihalos ($10^5 < M_{\rm h} < 10^6 \, M_\odot$) and the impact of individual stellar feedback on the chemodynamical evolution of their environment.  However, the high computational costs of modeling individual stars, tracking yields for multiple heavy elements, and tracing radiative feedback, all at high resolution, make developing a cosmological model to study the effects of Pop III stars on present-day observations difficult.  High computational costs also prevent thorough exploration of parameter space for assumed values and functions such as the IMF, SN injection energy, DM streaming velocities, etc., all of which impact the contribution of Pop III to the chemodynamical evolution of galaxies.

Semi-analytic models (SAMs) are computationally efficient alternatives to full hydrodynamic cosmological simulations, running orders of magnitude faster.  By combining DM halo merger trees with analytic prescriptions for star formation, radiation, feedback, metal enrichment and so forth, SAMs can be used to rapidly explore parameter space to study the effects of altering one or more of the many tunable parameters in the models. Importantly, SAMs also enable studies on the larger domain sizes needed to statistically capture the stochastic variation in star formation, metal enrichment, and reionization expected in the early Universe.   Although many SAMs do not implement modeling for the first stars, instead initializing a metallicity floor representative of their enrichment, there are now numerous SAMs that have been used to study Pop III and the first galaxies \citep[e.g.][]{Trenti2009,Komiya2010,Crosby2013,LiuBromm2020,Visbal2020,ASLOTH,Corazza2022,HegdeFurlanetto2023,Feathers2024,Ventura2024,Ishiyama2025}.  Ultimately, SAMs with Pop III models, run and calibrated to present day observations, are critical to the exploration of broad parameter spaces to interpret the observational signatures from Pop III and their descendants.

Calibrating a SAM for Pop III star formation and the transition to Pop II requires simulations that resolve both the first Pop III–forming halos and those that subsequently host Pop II. Because only a few Pop III stars form per halo, they must be modeled as individual stars drawn from an IMF rather than as stellar population particles; otherwise, stellar masses converge toward the particle mass resolution \citep{Brauer2025b}. Additionally, the feedback from individual stars is weaker and less bursty than that from particles containing single stellar populations \citep{Smith2021}, whose use can yield higher stellar masses at a fixed halo mass \citep{JeonKo2024,Brauer2025b}. Thus, calibrating a SAM to simulations that track individual Pop III stars is essential to capture the evolution of both primordial and early metal-enriched environments.

In this paper, we calibrate a SAM \citep{Visbal2020} to the \Aeos simulations \citep{AeosMethods} in order to develop a computationally efficient model of Pop III star formation and galaxy evolution in minihalos.  We summarize the relevant details from the \Aeos simulations and the \citet{Visbal2020} SAM in Section \ref{sec:sims}. Our method for calibrating the SAM to \Aeos is described in Section \ref{sec:calib}, with an analysis of the resulting Pop III and Pop II star formation in the SAM in Section \ref{sec:results}.  We then apply the calibrated SAM to a selection of other cosmological volumes and compare the predicted number density of active Pop III sources to observations in Section \ref{sec:application}, and conclude in Section \ref{sec:conclusions}.

\section{Computational Methods} \label{sec:sims}
We summarize here the relevant physics and model components used in the \Aeos simulations \citep{AeosMethods} and the Renaissance-calibrated SAM \citep{Visbal2020,Hazlett2025}, and refer the reader to the respective methods papers for greater detail.

\subsection{AEOS}
\Aeos \citep{AeosMethods} is a 1 comoving Mpc per side cosmological hydrodynamics simulation run with the adaptive mesh code {\sc Enzo} \citep{Enzo2014,Enzo2019} that uses the best fit cosmological parameters from \citet{Planck2014}.  Star particles in \Aeos represent individual massive stars that contribute feedback on the time scale of the simulation, enabling the study of the effects of individual stellar feedback. With a DM mass resolution of 1840 M$_\odot$, and a maximum spatial resolution of 1 proper pc, \Aeos can resolve minihalos (${\rm M}_{\rm vir} \sim 10^5 \,{\rm M}_\odot$), which are thought to be the formation sites of the first Pop III stars.  We limit our summary of \Aeos here to the details relevant for the calibration of the SAM.

\subsubsection{Radiative Cooling} \label{subsubsec:rad_cooling}
\Aeos uses a modified version of {\sc grackle} \citep{GrackleMethod} to follow radiative heating and cooling from a \citet{HM2012} UV background that is extended to high redshifts and metal lines.  Additionally, \Aeos includes background Lyman-Werner (LW) contributions adopted from \citet{Emerick2019} and \citet{Wise2012a}, with a high redshift contribution adopted from \citet{Qin2020}.

\subsubsection{Star Formation} \label{subsubsec:star_form}
Stars in the simulation form stochastically from an IMF when at least $100 \, M_\odot$ of gas in a given cell satisfy the following star-formation criteria: (i) converging gas flow ($\nabla \cdot \mathbf{v} < 0$), (ii) $T < 500 \, \rm K$, and (iii) $n > 10^4 \, \rm cm^{-3}$.

\paragraph{Population III}
In addition to the above constraints, gas that has a minimum molecular hydrogen fraction $f_{\rm H_2} > 0.0005$ \citep{Susa2014} and metallicity $Z < 10^{-5} \,\rm Z_\odot$ goes on to form Pop III stars.  Following \citet{Wise2012a}, Population III stellar masses are drawn from a \citet{Salpeter1955} IMF with a power-law slope $\alpha=2.3$ above a characteristic mass $M_{\rm char}$ and an exponential cutoff below.  Lifetimes for these stars are adopted from \citet{Schaerer2002}.  For our calibration, we use the {\sc Aeos10} simulation, which uses $M_{\rm char} = 10 \, M_\odot$ and a mass range of $1 < M_*<100 \, M_\odot$.

\paragraph{Population II}
Pop II stellar masses are drawn from a \citet{Kroupa2001} IMF with a range of $0.08$--$120 \, M_\odot$.  Pop II stars are treated as individual particles if their mass exceeds $2 \, M_{\odot}$. The remaining stars are agglomerated into single particles as they provide neither feedback nor chemical enrichment during the simulation. Lifetimes and asymptotic giant branch (AGB) phases are set using PARSEC \citep{Bressan2012}.

\subsubsection{Stellar Feedback}
\Aeos implements both AGB and massive stellar winds, core-collapse SNe, Type Ia SNe, and ionizing radiation from massive stars.  Mass loss rates from winds are assumed to be constant over the relevant ejection phase.  All SNe explode with a fixed energy of $10^{51}$ erg, with core-collapse SNe occurring for Pop III stars with $10 \, M_\odot < M_* < 100 \, M_\odot$ and Pop II stars with $8 \, M_\odot < M_* < 25 \, M_\odot$, and Type Ia SNe occuring for Pop II stars with $3 \, M_\odot < M_* < 8 \, M_\odot$.  Ionizing radiation from Pop III stars use luminosities from \citet{HegerWoosley2010} and lifetimes from \citet{Schaerer2002}, while Pop II luminosities are taken from OSTAR2002 \citep{Lanz2003} when available, otherwise a scaled blackbody spectrum.

\subsection{Semi-analytic Model calibrated to Renaissance}

We use an updated version of the SAM presented in \citet{Visbal2020}, which takes DM halo merger trees with 3D spatial coordinates from cosmological $N$-body simulations and applies analytic prescriptions for Pop III and metal-enriched star formation. The model is calibrated to the Renaissance simulations \citep{2015oshea}, cosmological hydrodynamics simulations performed with the adaptive mesh refinement code \textsc{Enzo} \citep{2014Bryan,Enzo2019} that model the formation of the first stars and galaxies. The SAM also includes feedback from LW radiation and the expansion of H~\textsc{ii} regions, computed on a grid using fast Fourier transforms, along with external metal enrichment.

Whether a halo forms Pop III stars or not is a key parameter in the model. With a DM particle mass of $2.9\times10^{4} \ M_\odot$, halos with $M_{\rm h} \lesssim 2\times10^{6} \ M_\odot$, including $10^{5-6} \ M_\odot$ minihalos where Pop III stars are thought to form in the early Universe, are poorly resolved \citep{2015oshea}. This limits the robustness of the Pop III calibration using Renaissance. In contrast, \Aeos resolves a large sample of such minihalos, allowing a more reliable calibration that can be applied over a broader range of halo mass.

The Renaissance-calibrated SAM described by \citet{Hazlett2025} reproduces both Pop III and metal-enriched SFRDs within a factor of $\sim$2 compared to the calibrating simulation between $11 < z < 25$ in a representative cosmological volume. For regions that are more overdense or rarefied, we found metal-enriched SFRDs that agreed with Renaissance within a factor of $\sim 2$ at high-redshift but differ by a factor of $\sim$3 or more at later times.

\section{Calibration to \Aeos} \label{sec:calib}
To calibrate the SAM to \Aeos, we identify halos that host the earliest Pop III star formation in \Aeos\ and track their subsequent metal-enriched star formation. We determine the critical halo mass for Pop III formation and the stellar masses of Pop III stars, introducing a key modification to $M_{\mathrm{crit}}$ that better reproduces the distribution of halos that form Pop III stars in \Aeos. We measure the delay times before the onset of Pop II star formation and parameterize metal-enriched Pop II starbursts in low-mass halos using power-law fits, combining these with the Renaissance-calibrated bursts for higher-mass halos. Finally, we implement steady star formation and feedback processes, including H~\textsc{ii} bubble growth and external metal enrichment, to produce a self-consistent model that reproduces the timing, mass, and distribution of Pop III and Pop II star formation observed in \Aeos.

\subsection{Halo Sample}

We identify DM halos using the halo finder \textsc{rockstar} \citep{2013behroozi_a} and calculate halo merger trees with \textsc{consistent trees} \citep{2013behroozi_b}. For our calibration we focus on the 216 halos in \Aeos that host the first instances of Pop III star formation along with the stellar mass histories for $\sim 20$ halos that experience subsequent metal-enriched star formation.

\subsection{Star Formation Parametrization} \label{subsec:star_param}

Our model tracks Pop III star formation in pristine halos and the subsequent onset of metal-enriched stars that occurs after a delay period. Metal-enriched star formation then proceeds through a bursty phase of rapid starbursts separated by quiescent periods, followed by a steady phase once halos reach a critical virial temperature allowing sustained star formation. This critical temperature corresponds to the halo mass below which photoheating from reionization suppresses star formation \citep[e.g.][]{Shapiro1994,Thoul1996,Gnedin1998,Gnedin2000,Dijkstra2004,Hoeft2006,Okamoto2008,Sobacchi2013,Noh2014}. Rather than modeling individual stars, the SAM calculates the total stellar mass formed in each halo over time intervals set by the DM merger trees.

It is important to note that our star formation prescription differs from the calibration presented in \citet{Hazlett2025} in several key ways. We calibrate Pop III star formation in our model exclusively to \Aeos and exclude prior calibration to the Renaissance simulations. This is due to the improved resolution of the minihalos hosting Pop III formation in \Aeos. We also calibrate to metal-enriched star formation, extending the metal-enriched burst mass relation from Renaissance to lower halo masses. We discuss details for our star formation prescription calibrated to \Aeos in the following section.

\subsubsection{Population III} \label{subsubsec:popIII}

Pop III stars form when pristine gas collapses to high densities, enabled by H$_2$ and atomic cooling. This is parameterized by $M_{\mathrm{crit}}$, the critical halo mass for Pop III star formation to occur, which strongly influences early star formation rates in the SAM.

The value of $M_{\mathrm{crit}}$ depends on several processes. LW radiation dissociates H$_2$, raising $M_{\mathrm{crit}}$ in regions with strong LW backgrounds \citep{Dekel1987,Haiman1997,Haiman2000,Machacek2001,Wise2007,Oshea2008,Ahn2009,Visbal2014}, although self-shielding in minihalos can partially offset this effect. In addition, the baryon–DM streaming velocity $v_{\rm bc}$ suppresses gas accretion, further increasing $M_{\mathrm{crit}}$ in high-velocity regions \citep{Kulkarni2021,Schauer2021,Nebrin2023}.

For our model, we compute the critical mass using a modified version of the fitting function found in \citet{Kulkarni2021} that incorporates H$_2$ self-shielding along with a dependence on the LW background flux $J_{\rm LW}$ and the baryon-DM streaming velocity:
\begin{equation}
     M_{\rm crit} = M_{z = 20}(J_{\rm LW}, v_{\rm bc})\bigg(\frac{1+z}{21}\bigg)^{-\alpha(J_{\rm LW}, v_{\rm bc})}
     \label{eq:MCrit}
\end{equation}
where 
\begin{equation}\begin{split}
    M_{z=20}(J_{\rm LW}, v_{\rm bc}) & = 2.25 \times (M_{z = 20})_0(1 + J_{\rm LW}/J_0)^{\beta_1} \\
    & \times (1 + v_{\rm bc} / v_0)^{\beta_2}(1 + J_{\rm LW}v_{\rm bc}/Jv_0)^{\beta_3}
\end{split}\end{equation}
and
\begin{equation}\begin{split}
    \alpha(J_{\rm LW}, v_{\rm bc}) & = \alpha_0(1+J_{\rm LW}/J_0)^{\gamma_1}(1+v_{\rm bc}/v_0)^{\gamma_2} \\ 
    & \times (1+J_{\rm LW}v_{\rm bc}/J_0v_0)^{\gamma_3} - 0.8. 
\end{split}    
\end{equation}
The constants in these functions are given in \citet{Kulkarni2021}.

To align with the physical conditions in \Aeos, we employ a $J_{\mathrm{LW}}$ equivalent to the LW background described in Section \ref{subsubsec:rad_cooling} and $v_{\mathrm{bc}} = 0 \ \text{km s}^{-1}$. To produce a similar distribution of halos that form Pop III stars, we modified the fit from \citet{Kulkarni2021} by multiplying $M_{\mathrm{z=20}}$ by 2.25 and reducing $\alpha$ by 0.8 to increase $M_{\mathrm{crit}}$ and flatten the slope at high redshift (Equation \ref{eq:MCrit}). Additionally, we implement a logarithmic normal scatter to $M_{\mathrm{crit}}$ with a standard deviation of $\log_{10}\left( M_{\mathrm{crit}} / M_{\odot} \right) = 0.4$. This scatter is assigned at halo formation and passed along to descendant halos. If there is a merger between halos, a random standard deviation from from one of the halos is assigned to the descendant.

\begin{figure}
\epsscale{1.1}
\plotone{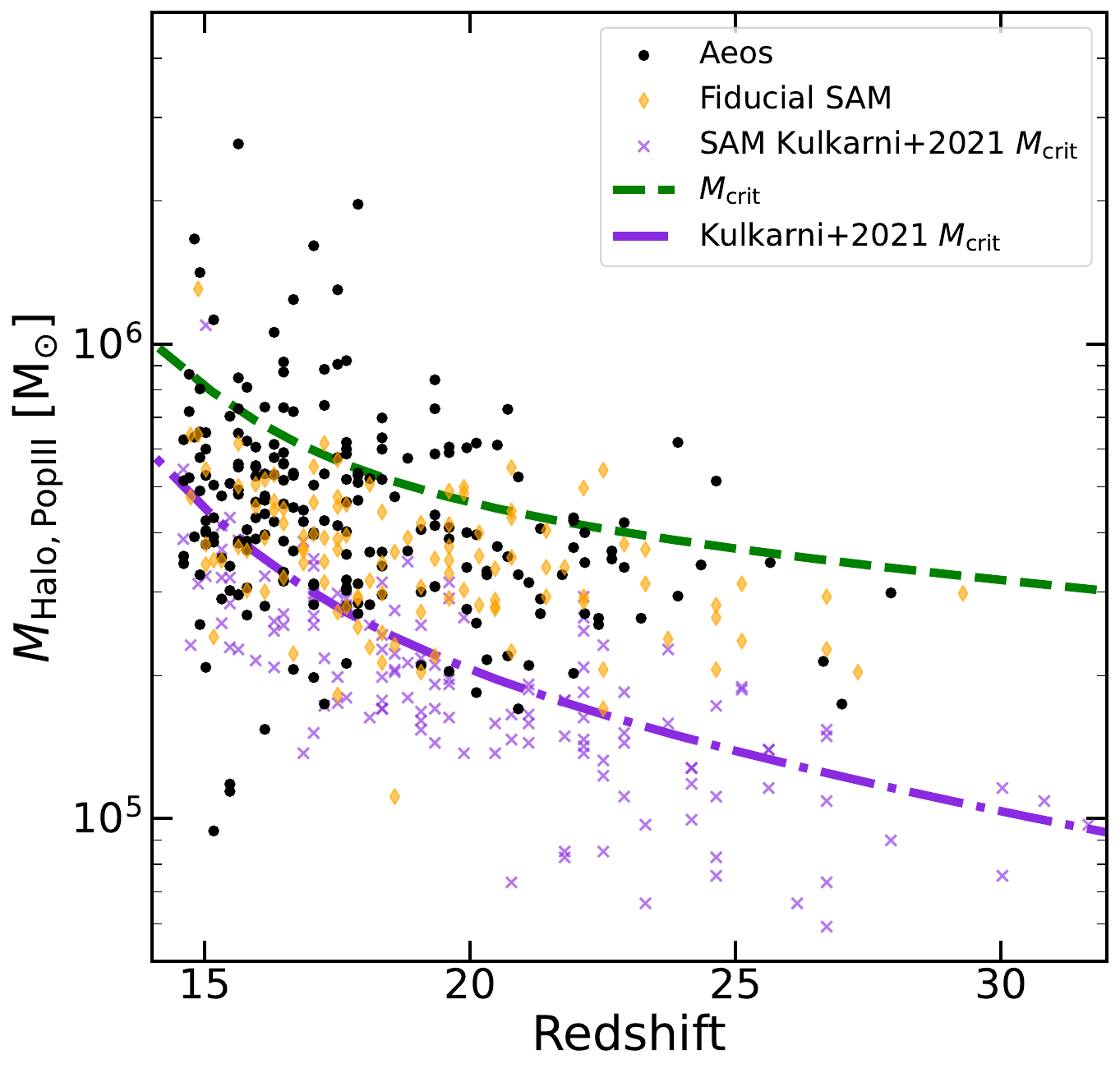}
\caption{Virial masses for halos hosting the first Pop III stars in \Aeos {\em (black points)} compared to a single realization of our semi-analytic model {\em (orange diamonds)} and a realization of the SAM using an unmodified $M_{\mathrm{crit}}$ from \citet{Kulkarni2021}, shown with {\em purple crosses}. The {\em green dashed line} shows our modified $M_{\mathrm{crit}}$ from Equation (\ref{eq:MCrit}), while the {\em purple dash-dotted} line shows the unmodified $M_{\mathrm{crit}}$ from \citet{Kulkarni2021}. Our modification corrects for the effect introduced by applying scatter to $M_{\mathrm{crit}}$, which otherwise produces an artificial excess of low mass Pop III forming halos. This adjustment brings the distribution of Pop III host halos into closer agreement with \Aeos. \label{fig:first_popIII}}
\end{figure}

Figure~\ref{fig:first_popIII} shows the original and modified functions. This modification accounts for how the scatter in $M_{\mathrm{crit}}$ interacts with the halo mass function in merger tree realizations. While the original fitting function of \citet{Kulkarni2021} defines $M_{\mathrm{crit}}$ as the mass where 50\% of halos contain sufficient cold, dense gas, applying this definition directly leads to a bias toward lower mass halos in our SAM. Because there are many more halos just below $M_{\mathrm{crit}}$ than above it, the introduction of scatter causes a systematic excess of low-mass Pop III forming halos compared to \Aeos. By increasing $M_{\mathrm{z=20}}$ and flattening the redshift dependence, our modified $M_{\mathrm{crit}}$ better represents the subset of halos that actually cross the threshold for Pop III star formation in \Aeos, while remaining broadly consistent with the $v_{\mathrm{bc}}=0$ and $0<J_{\mathrm{21}}<1$ results shown in Figure 3 of \citet{Kulkarni2021}.

Figure \ref{fig:first_popIII} shows the distribution of DM halo masses hosting Pop III star formation as a function of redshift. Our modifications to Equation \ref{eq:MCrit} results in our SAM producing both a similar number and overall distribution of Pop III forming halos to \Aeos.
This modification to $M_{\mathrm{crit}}$ is a key result of our work, as it ensures that the SAM accurately reproduces the number and distribution of Pop III-forming halos seen in \Aeos. Taking into account how the scatter in $M_{\mathrm{crit}}$ interacts with the halo mass function, the adjustment prevents the systematic overproduction of low-mass Pop III halos shown with the purple crosses in Figure \ref{fig:first_popIII}. Any SAM that aims to model early star formation across a realistic halo mass range should incorporate this modified $M_{\mathrm{crit}}$ to reliably capture the timing and sites of Pop III star formation.

To determine the overall minimum mass for Pop III star formation, we take the minimum of $M_{\mathrm{crit}}$ and the atomic cooling threshold found in hydrodynamical cosmological simulations \citep{Fernandez2014}
\begin{equation} M_{\mathrm{a}} = 5.4 \times 10^{7} \left( \frac{1+z}{11} \right)^{-3/2} M_{\mathrm{\odot}}.
\end{equation}

We calibrate Pop III star formation in our model to produce stellar masses consistent with \Aeos. First, we determine the mass of Pop III stars formed in a halo after its mass exceeds $M_{\mathrm{crit}}$ by sampling from a normalized cumulative distribution function
\begin{equation}
     F(x) = 2.29 e^{-1.46 x} + 0.06 e^{-0.24 x},
     \label{eq:CDF}
\end{equation}
where $x$ is the number of star forming cells in a Pop III star formation event.  Because \Aeos forms stars in a given cell once the gas mass available for star formation exceeds $100 M_\odot$, this results in a build-up of masses at multiples of $100M_\odot$ for a given star formation event.  Thus, we parameterize the mass of stars that form by the number of star-forming cells per star formation event in \Aeos. In the SAM, we sample a number of star-forming cells from Equation \ref{eq:CDF}, and then for each Pop III star formed up to or slightly exceeding $\sim 100 M_\odot$, we assign a stellar mass by sampling from the same Salpeter IMF as \Aeos, as described in Section \ref{subsubsec:star_form}.

\subsubsection{Population II} \label{subsubsec:popII}

Following enrichment from Pop III stars, metals are dispersed into the nearby ISM and intergalactic medium, fueling the formation of Pop II stars. If a halo is in a bubble that exceeds our fiducial critical metallicity of $Z_{\mathrm{crit}} = 10^{-5} \ Z_{\mathrm{\odot}}$, Pop II star formation can proceed. Compared to radiative feedback, metals take comparatively longer to propagate throughout the host halo and to nearby halos. To model the time it takes for the gas to recover from feedback effects following the birth of the first Pop III stars in a halo, we introduce a delay time, $t_{\mathrm{delay}}$, before the beginning of Pop II star formation.

\begin{figure}
\epsscale{1.1}
\plotone{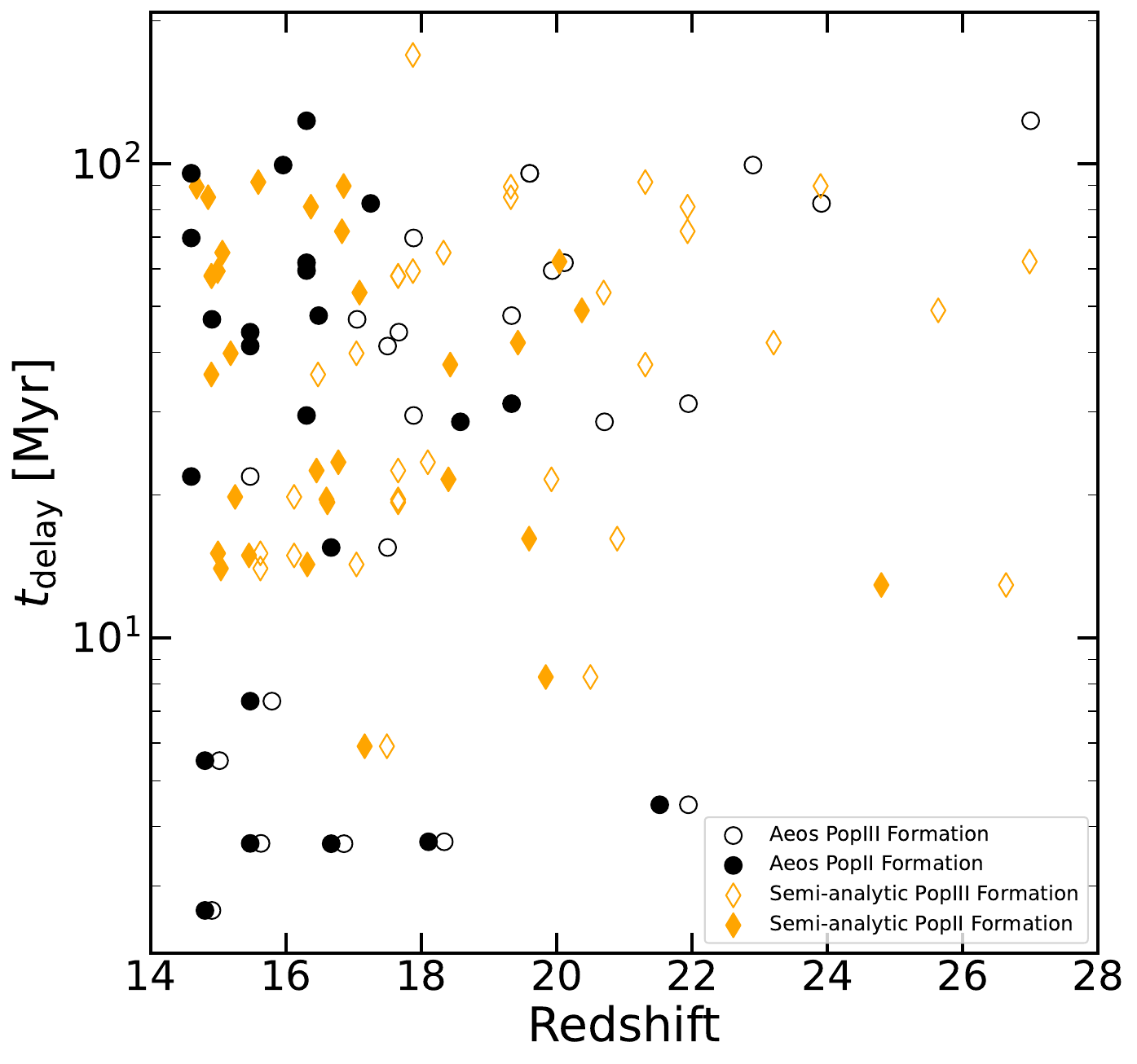}
\caption{The delay period before the formation of metal-enriched Pop II stars following the death of the first Pop III star for \Aeos {\em (black)} and a single realization of the SAM {\em (orange)}. Open points are the Pop III formation redshift and filled points are the redshift of Pop II formation for a single halo.
\label{fig:popIII_to_popII_delay}}
\end{figure}

In Figure \ref{fig:popIII_to_popII_delay}, we show the distribution of $t_{\mathrm{delay}}$ for \Aeos halos compared to a single realization of the SAM. We find the best agreement with \Aeos using a logarithmic normal distribution centered on $t_{\mathrm{delay}} \approx 300$ Myr with a standard deviation of log$_{\mathrm{10}}(t_{\mathrm{delay}}/1\,\mathrm{Myr}) = 0.65$. We also tested setting $t_{\mathrm{delay}}$ to a constant value of 100 Myr and found that our SAM produced roughly an order of magnitude more Pop II hosting halos than \Aeos for $z < 25$, along with many orders of magnitude higher total Pop II stellar mass between $20 < z < 25$. This strong sensitivity to $t_{\mathrm{delay}}$ demonstrates that it is a key parameter governing the overall star formation history. The underlying physics that determines $t_{\mathrm{delay}}$ remains uncertain, but could be related to: how radiative and supernova feedback couple to the gas \citep{Jeon2014}, how efficiently metals mix within halos, and how rapidly gas can be re-accreted to form subsequent generations of stars.

It is important to note that many halos that form Pop III stars in \Aeos are expected to later host Pop II formation. However, because the final snapshot of \Aeos occurs at $z=14.6$, the simulation ends before this transition can be fully realized. Although our prescription for $t_{\mathrm{delay}}$ reproduces the approximate timing and number of halos that undergo Pop II formation, the actual complete $t_{\mathrm{delay}}$ distribution remains uncertain. This transition cannot be fully captured in simulations like \Aeos that conclude midway through reionization due to computational limitations, and this inability to resolve the true $t_{\mathrm{delay}}$ distribution remains an inherent limitation of our calibration framework.

By the final snapshot at $z = 14.6$, most halos in \Aeos that have experienced internal metal enrichment are actively forming stars in the bursty stage of Pop II star formation. In these halos, we assume that once the delay time $t_{\mathrm{delay}}$ has elapsed after Pop III formation, sufficient metals are present to trigger Pop II formation. In contrast, for externally enriched halos, Pop II formation depends explicitly on the metallicity of the halo exceeding the critical threshold, as described in the following subsection.

\begin{figure*}
\epsscale{1.0}
\plotone{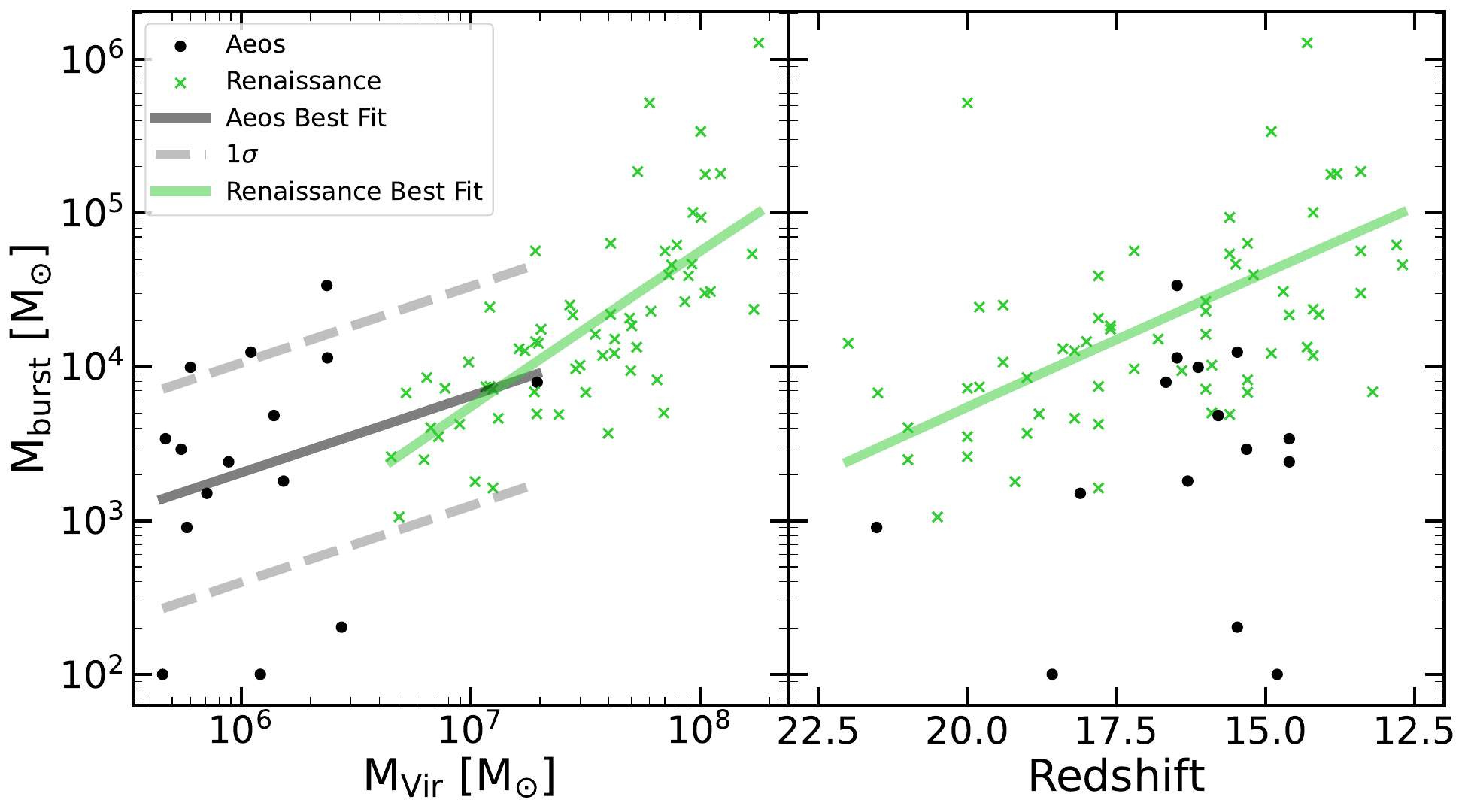}
\caption{Metal-enriched Pop II starburst masses for the \Aeos halos {\em (black points)} and the sample of Renaissance halos {\em (green crosses)} from \citet{Hazlett2025} during the bursty stage compared with the halo virial mass and redshift at the starburst. The best fit power law {\em (solid)} along with 1$\sigma$ lines {\em (dashed)} is shown. Overall, the \Aeos halos extend the burst mass relation to lower mass halos than Renaissance.
\label{fig:burst_mass}}
\end{figure*}

We evaluated multiparameter power-law fits to the halo virial mass and redshift at the onset of the starburst using the \textsc{scipy} curve-fit package and applied them to the \Aeos distribution. We find that the Pop II stellar mass formed in a burst is correlated with the halo virial mass and increases less steeply than in the Renaissance halos, as shown in Figure \ref{fig:burst_mass}. Unlike the Renaissance sample, we find only a weak correlation between starburst redshift and stellar mass, likely due to the small number of Pop II-forming halos in the \Aeos volume and that most of the starbursts occur at a similar redshift. We parameterize the metal-enriched stellar mass formed in a burst as:

\begin{equation}
M_{\mathrm{burst,aeos}} = \alpha_{\mathrm{aeos}} \left( \frac{M_{\mathrm{vir}}}{10^{6} M_{\mathrm{\odot}}} \right)^{\gamma_{\mathrm{aeos}}} M_{\mathrm{\odot}},
\label{eqn:Mburst_eqn}
\end{equation}

where $M_{\mathrm{vir}}$ is the virial mass of the halo. The best-fit parameters are $\alpha_{\mathrm{aeos}} = 2.05 \times 10^{3}$ and $\gamma_{\mathrm{aeos}} = 0.5$. We find that the distribution of the sample around $M_{\mathrm{burst,aeos}}$ is logarithmic normal with a standard deviation of log$_{\mathrm{10}}\left( M_{\mathrm{burst,aeos}} / M_{\mathrm{\odot}} \right) = 0.71$. As halos reach larger masses at lower redshifts, the stellar mass of the Pop II starburst increases. In general, the distribution and best fit for $M_{\mathrm{burst,aeos}}$ are consistent with the Renaissance simulations and extend the prescription introduced in \citet{Hazlett2025} to lower mass halos that could not be resolved.

To combine the Pop II star formation calibrations in \Aeos and Renaissance, we implement a halo mass threshold separating the \Aeos calibrated Pop II bursts from the Renaissance calibrated bursts in higher mass halos. Based on the distribution of \Aeos and Renaissance starbursts in Figure \ref{fig:burst_mass}, halos in the bursty stage with masses $M_{\rm vir} < 4 \times 10^{6}\, M_{\mathrm{\odot}}$ will have Pop II burst masses set by Equation (\ref{eqn:Mburst_eqn}). Halos with masses $M_{\rm vir} > 4 \times 10^{6} \,M_{\mathrm{\odot}}$ instead use the Renaissance-calibrated Pop II burst masses that scale more steeply with the halo mass and redshift. This threshold reflects the physical transition between shallow and deep potential wells: low-mass halos are more susceptible to supernova-driven gas loss, limiting the total stellar mass that can form per burst, whereas higher-mass halos can retain a larger fraction of their gas and sustain more efficient, repeated starbursts. Any halo in the bursty stage will have shorter quiescent periods between bursts as it grows more massive, consistent with the fits presented in \citet{Hazlett2025} and with the expectation that deeper potentials shorten recovery times following SN feedback.

Halos undergoing Pop II star formation will then proceed to the steady star formation stage, which occurs once the halos exceed a critical virial temperature of about $2 \times 10^4$~K. In this regime, star formation is well described by a bathtub-type model where gas inflow, star formation, and SN-driven outflows regulate the system. Our calibration to Renaissance favors an efficiency of about 1\% of gas converted into stars per freefall time, with some scatter from halo to halo. The steady stage reproduces the smoother star formation seen in more massive halos, while still accounting for the variability introduced by feedback. More details about the Renaissance-calibrated star formation model can be found in Table \ref{tab:overview_table} and in \citet{Hazlett2025}.

\begin{deluxetable*}{lcccc}
\tablecaption{Parameters for Star Formation in the Semi-analytic Model \label{tab:overview_table}}
\tablehead{
\colhead{Parameter} & \colhead{Description} & \colhead{Fiducial Equation} & \colhead{Fit Values} & \colhead{Standard Deviation}
} 
\startdata 
$t_{\mathrm{delay}}$ & Delay before Pop II star formation & - & $300$ Myr & 0.65\\
\hline
\rule{0pt}{4ex}
$M_{\mathrm{burst,aeos}}$ & \Aeos Pop II starburst mass & $\alpha_{\mathrm{aeos}} \left( \frac{M_{\mathrm{vir}}}{10^{6} M_{\mathrm{\odot}}} \right)^{\gamma_{\mathrm{aeos}}} M_{\mathrm{\odot}}$ & $\alpha_{\mathrm{aeos}} = 2.05 \times 10^{3}$ & $0.71$\\
{       } & {       } & {       } & $\gamma_{\mathrm{aeos}} = 0.5$ & {       }\\
\hline
\rule{0pt}{4ex}
$M_{\mathrm{burst,Ren}}$ & Renaissance Pop II starburst mass & $\alpha_{\mathrm{Ren}} \left( \frac{M_{\mathrm{vir}}}{10^{7} M_{\mathrm{\odot}}} \right)^{\gamma_{\mathrm{Ren}}} \left( \frac{1 + z}{20} \right)^{\delta_{\mathrm{Ren}}} M_{\mathrm{\odot}}$ & $\alpha_{\mathrm{Ren}} = 5.36 \times 10^{3}$ & $0.41$\\
{       } & {       } & {       } & $\gamma_{\mathrm{Ren}} = 1.09$ & {       }\\
{       } & {       } & {       } & $\delta_{\mathrm{Ren}} = 0.58$ & {       }\\
\hline
\rule{0pt}{4ex}
$t_{\mathrm{quies}}$ & Quiescent period after Pop II starburst & $\epsilon \left( \frac{M_{\mathrm{vir}}}{10^{7} M_{\mathrm{\odot}}} \right)^{\kappa} \text{Myr}$ & $\epsilon = 44.8$ & $0.26$\\
{       } & {       } & {       } & $\kappa = -0.39$ & {       }\\
\hline
\rule{0pt}{4ex}
$f_{\mathrm{II}}$ & Steady stage star formation efficiency & $\dot{M}_{\mathrm{*}} = \frac{f_{\mathrm{II}}}{t_{\mathrm{ff}}} M_{\mathrm{gas}},$ & $f_{\mathrm{II}} = 0.01$ & $0.42$\\
{       } & {       } & $\dot{M}_{\mathrm{gas}} = \dot{M}_{\mathrm{acc}} - \dot{M}_{\mathrm{*}} - \eta_{\mathrm{SN}} \dot{M}_{\mathrm{*}}$ & {       } & {       }\\
\enddata 
\tablecomments{$t_{\mathrm{delay}}$, $M_{\mathrm{burst,aeos}}$, $M_{\mathrm{burst,Ren}}$, $t_{\mathrm{quies}}$, $f_{\mathrm{II}}$ each follow a logarithmic normal distribution. The last column shows the standard deviation of the log$_{\mathrm{10}}$ of the corresponding quantity mentioned in the text. Halos in the bursty stage with masses $< 4 \times 10^{6} M_{\mathrm{\odot}}$ will have Pop II burst masses set by $M_{\mathrm{burst,aeos}}$ and halos with masses $\geq 4 \times 10^{6} M_{\mathrm{\odot}}$ instead use $M_{\mathrm{burst,Ren}}$. Halos with masses $>10^{9} M_{\mathrm{\odot}}$ extrapolate beyond the calibration to the Renaissance simulations.}
\end{deluxetable*}

\subsection{H~\textsc{ii} Bubbles and External Metal Enrichment} \label{subsec:HII_bubbles_ExtMetal}

To include feedback effects from reionization and external metal enrichment, we implemented H \textsc{ii} and metal bubbles into the SAM. To determine the size and growth of the ionized H\textsc{ii} bubbles we use the fast Fourier transform method introduced in \cite{Visbal2020}. We find the total number density of ionizing photons produced in dark-matter halos that have escaped to the surrounding intergalactic medium for each cell in a $256^{3}$ cubic grid. This is set by the total mass of stars ever formed in each cell along with the escape fraction of hydrogen ionizing photons from halos hosting Pop III and metal-enriched stars, $f_{\mathrm{esc,III}}$ and $f_{\mathrm{esc,II}}$, respectively. A cell is considered ionized if it exceeds a threshold dependent on the cosmic mean comoving density of hydrogen atoms. \citet{Behling2025} found SAMs that do not account for self-shielded neutral gas in dense filaments overestimate the fraction of halos in H\textsc{ii} regions impacted by reionization by up to an order of magnitude. Therefore, it is likely that our calibrated model overestimates the reionization feedback.

Since \Aeos only covers the early stages of reionization at redshifts of $z>14.6$, the extent of ionized H~\textsc{ii} bubbles remain fairly localized, only impacting Pop III star formation in a handful of halos. Using escape fractions of ionizing radiation of $f_{\mathrm{esc,II}} = 0.02$ for Pop II stars and $f_{\mathrm{esc,III}} = 0.1$ for Pop III stars, we obtain an ionization fraction of $\sim 0.05$ at $z=15$ compared to the \Aeos value of $\sim 0.08$ \citep{Brauer2025b}. In our model, if a halo with no prior star formation is located in an ionized region, we increase the critical mass for Pop III formation to 

\begin{equation}M_{\mathrm{crit}} = 1.5 \times 10^{8} \left( \frac{1+z}{11} \right)^{-3/2}M_{\mathrm{\odot}},
\end{equation}

consistent with \citep{Dijkstra2004}.

We also include the external enrichment of pristine halos by SN wind-driven metal bubbles that result in metal-enriched star formation \citep{Visbal2020}. If a halo is in a bubble that exceeds our fiducial critical metallicity of $Z_{\mathrm{crit}} = 10^{-5} \ Z_{\mathrm{\odot}}$, the Pop II star formation will proceed as described in Section \ref{subsubsec:popII}. Other works find a critical value near our fiducial choice \citep[e.g.][]{Bromm2003,Omukai2005,Smith2007,Smith2009}, but it could be significantly lower ($Z_{\mathrm{crit}} \approx 10^{-6} \ \text{Z}_{\mathrm{\odot}}$) if dust cooling is important, rather than just C and O \citep{Omukai2005,Jappsen2009}.

\section{Results} \label{sec:results}
The final snapshot of the \Aeos simulations is at $z=14.6$, while the DM merger trees used for the SAM extend to $z=8.3$. This allows the SAM to make predictions that go beyond the end of the \Aeos\ simulation, while maintaining the calibration to the \Aeos results. At $z=14.6$, \Aeos contains $\sim 200$ Pop III–forming halos and fewer than 20 Pop II-forming halos. The scarcity of Pop II sources has important implications for interpreting differences with the SAM, since stochastic effects dominate when the number of star-forming halos is small.

\begin{figure}
\epsscale{1.1}
\plotone{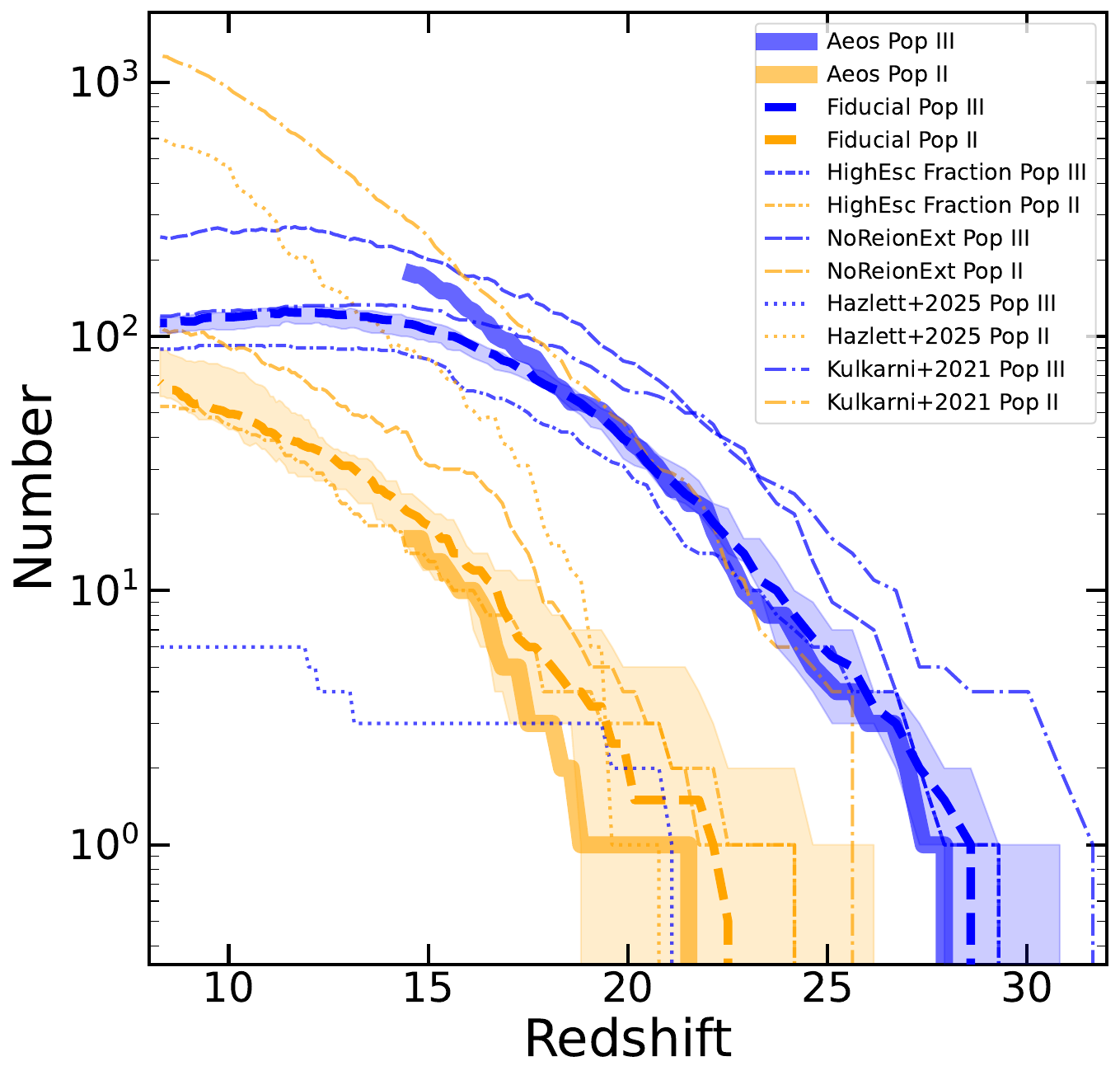}
\caption{Total number of halos hosting Pop III {\em (blue)} and Pop II stars {\em (yellow)}. The results from \Aeos are shown by {\em thick solid lines} for Pop III and Pop II, respectively. The results of 10 realizations of our fiducial SAM for Pop III and Pop II are also shown, with the {\em dashed line} representing the average and the {\em shaded region} extending between the maximum and minimum total stellar masses of the 10 simulations. The HighEsc {\em densely dot-dashed lines} show a realization using higher ionizing radiation escape fractions. The NoReionExt {\em densely dashed lines} show the impact of excluding reionization and external enrichment from the SAM. The Hazlett+2025 {\em dotted lines} show results from the SAM calibrated only to the Renaissance simulations. The Kulkarni+2021 {\em dot-dashed lines} show the impact of using the unmodified $M_{\mathrm{crit}}$ from \citet{Kulkarni2021}. \label{fig:total_halo_num}}
\end{figure}

\begin{figure}
\epsscale{1.1}
\plotone{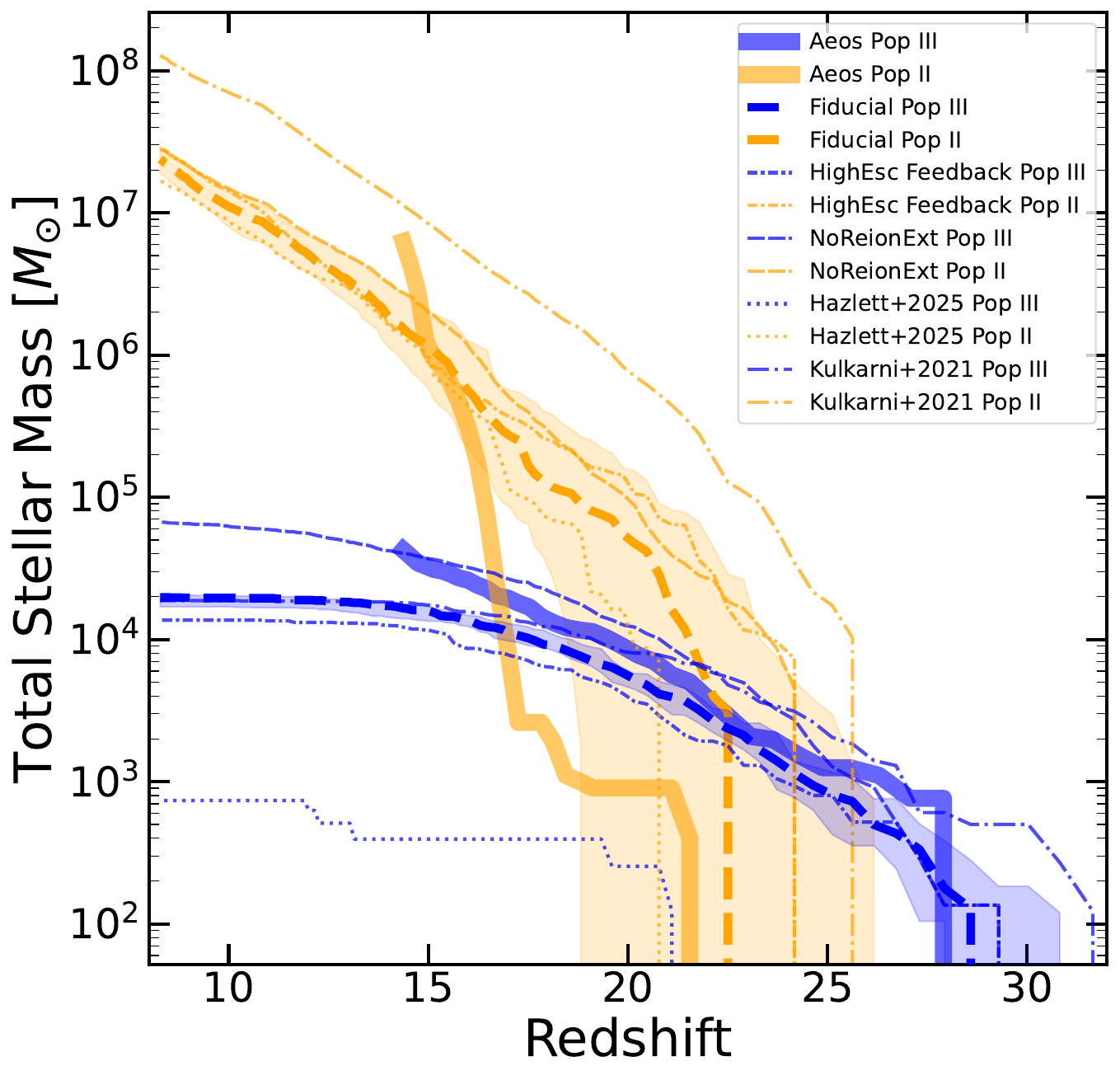}
\caption{Total stellar masses for Pop III and metal-enriched Pop II stars following the same convention for lines as the previous figure. \label{fig:total_stellar_mass}}
\end{figure}

 We run 10 independent SAM realizations with the fiducial physics summarized in Table \ref{tab:overview_table} and escape fractions of $f_{\mathrm{esc,III}} = 0.1$ and $f_{\mathrm{esc,II}} = 0.02$. Figure \ref{fig:total_halo_num} shows the total number of halos hosting Pop III and Pop II stars for the different random seeds, as well as varying physical assumptions. Figure \ref{fig:total_stellar_mass} shows the corresponding buildup in total stellar mass. For $z>17$, the fiducial models closely track the \Aeos results for both populations, demonstrating the strong consistency of our calibrated model with \Aeos. 

\begin{figure}
\epsscale{1.15}
\plotone{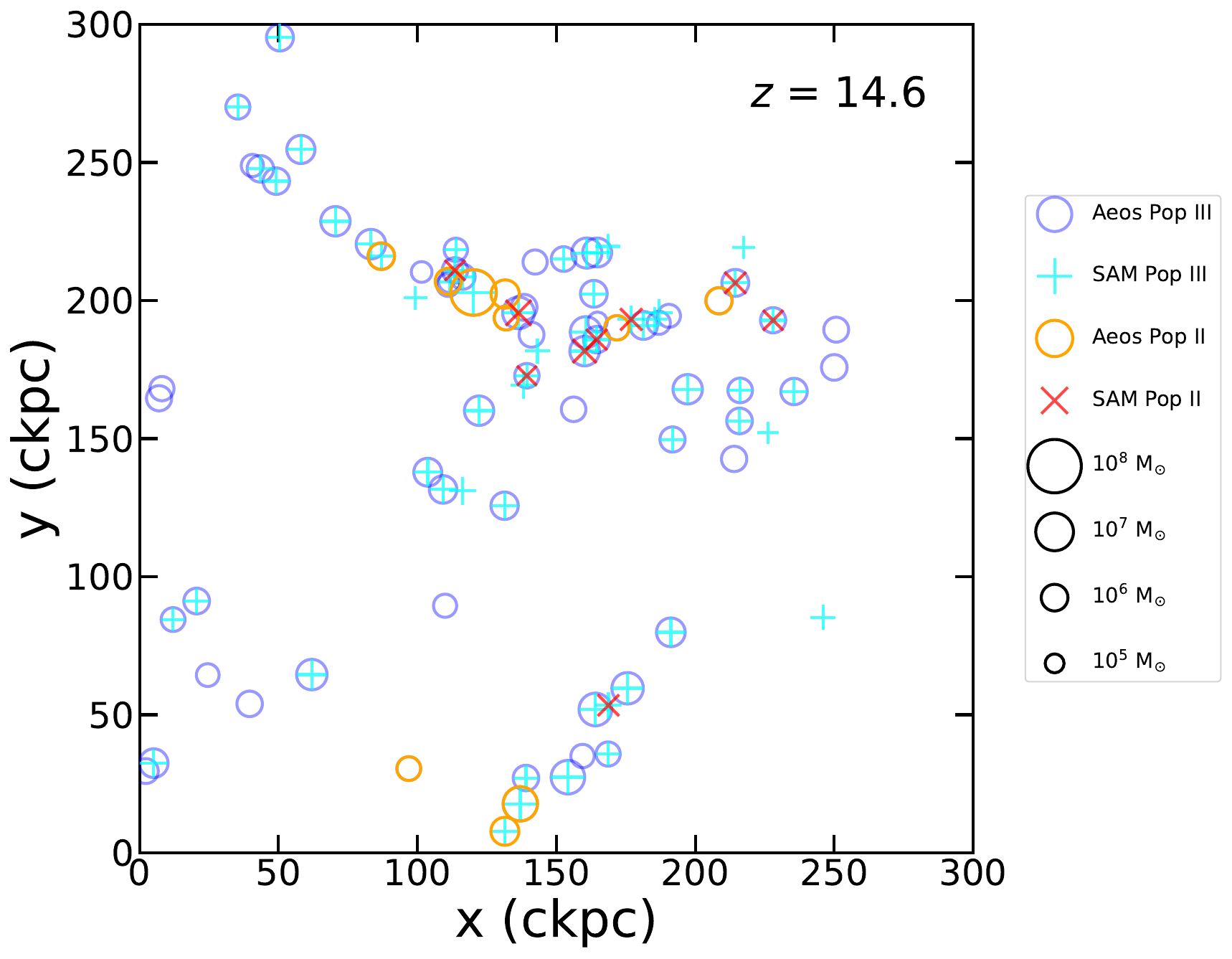}
\caption{A subregion of the \Aeos volume at the final simulation snapshot scaled in comoving kiloparsecs. Halos that have formed Pop III and Pop II stars in \Aeos are represented by {\em open blue and orange circles}. Halos that have formed Pop III and Pop II stars in the SAM are represented by {\em cyan and red crosses}. The size of the marker is scaled by the halo mass. Overlapping open circles and crosses are halos that have formed either Pop III or Pop II in both \Aeos and the SAM. Halos that have not formed stars are not shown. \label{fig:sf_halo_pos}}
\end{figure}

At $z<17$, the SAM begins to underestimate Pop III total stellar mass by a factor of $\sim$2 and the number of halos forming Pop III by $\sim$20\% by $z=14.6$ (Figure \ref{fig:total_halo_num}). These differences are likely related to increased local reionization effects in the SAM compared to \Aeos, which can delay Pop III formation in highly clustered regions. However, the halos that host Pop III stars are mostly consistent in both the SAM and \Aeos, as shown in Figure \ref{fig:sf_halo_pos}. In contrast, most of the halos hosting Pop II stars in the SAM form in different halos compared to \Aeos, reflecting a shortcoming in how $t_{\mathrm{delay}}$ is implemented. Because the actual distribution of $t_{\mathrm{delay}}$ is not fully captured in \Aeos, our model cannot employ a physically motivated calibration for this parameter, leading to discrepancies in which halos transition from Pop III to Pop II formation.

For Pop II, the number of star-forming halos matches well at $z=14.6$, although the total stellar mass of Pop II significantly diverges for $z>17$. This offset is likely driven by the fact that most Pop II star formation is concentrated in a handful of halos ($<20$ halos).  In particular, the delay time function is poorly sampled, and longer delay times in a few \Aeos halos over the SAM halos will magnify any discrepancy in star formation between the SAM and \Aeos.

Next, we test model variations to examine the sensitivity of the SAM to different physical conditions. Increasing the escape fractions to $f_{\mathrm{esc,III}}=0.5$ and $f_{\mathrm{esc,II}}=0.1$ (the HighEsc case) causes significant deviations from \Aeos: The ionization fraction approaches $\sim 0.17$ at $z=15$, the Pop III total stellar mass is underestimated by a factor of $\sim$4, and the number of Pop III halos by $\sim$2.5 at $z<20$. These discrepancies arise because larger HII regions increase the suppression of Pop III formation, shifting the impact of reionization to earlier times. Despite this, the total Pop II stellar mass and number of Pop II star forming halos remain roughly consistent with the fiducial runs, suggesting the earlier impact of reionization is likely efficient at suppressing star formation in low-mass halos.

Excluding both reionization and external metal enrichment (the NoReionExt case) leads to higher Pop III total stellar masses and halo counts between $14.6<z<25$ compared to the fiducial model, underscoring the regulating role of feedback processes over cosmic time. Pop III in \Aeos falls between the NoReionExt and the fiducial SAM results by $z=14.6$ likely due to overestimates of reionization feedback. However, by $z=14.6$ the results converge, once again underscoring the regulating role of feedback processes over cosmic time. In this case, the enhanced Pop III activity at earlier times also boosts Pop II formation for $z>17$.

Different behavior is seen in the Hazlett+2025 realization, calibrated only to the Renaissance simulations. Because the Renaissance volumes cannot resolve the $10^{5-6} \,M_{\mathrm{\odot}}$ minihalos thought to host Pop III star formation \citep{Kulkarni2021,Schauer2021,Nebrin2023}, an increase in $M_{\mathrm{crit}}$ near the resolution limit was introduced in that calibration. This suppresses early Pop III formation, delaying its onset and producing nearly two orders of magnitude fewer Pop III halos than \Aeos by $z=14.6$. Furthermore, the Renaissance-only calibration uses a broader distribution for the Pop III to Pop II delay time $t_{\mathrm{delay}}$, including delays much shorter than the $t_{\mathrm{delay}}$ distribution calibrated to \Aeos. This results in a more rapid transition to Pop II formation after the first Pop III stars compared to our fiducial SAM parameters.

Finally, the \citet{Kulkarni2021} realization illustrates the impact of our modified critical mass threshold $M_{\mathrm{crit}}$ (Section \ref{subsubsec:popII}). Without this correction, the SAM produces an artificial excess of low-mass Pop III halos, much earlier than the fiducial SAM and \Aeos. The use of the unmodified $M_{\mathrm{crit}}$ results in three times as many Pop III-forming halos as \Aeos at $z>17$, which in turn produces more numerous Pop II hosts earlier. For example, at $z = 20$ there are already $\sim70$ Pop II halos in the \citet{Kulkarni2021} realization of the SAM versus only 1--2 in \Aeos and the fiducial SAM. The total stellar mass predictions show smaller offsets in total Pop III mass which is consistent with \Aeos at $z>20$ and is about a factor of three lower at $z<20$, but the total Pop II mass is systematically overestimated.

Overall, our fiducial SAM realizations achieve excellent agreement with \Aeos across both stellar populations and redshifts, with only modest deviations attributable to small-scale reionization effects and the limited number of Pop II halos in \Aeos. The comparisons emphasize the necessity of our $M_{\mathrm{crit}}$ modification, which substantially improves consistency at high redshift, and demonstrates that the SAM provides a robust framework for extending predictions beyond the \Aeos volume to later cosmic times.

\section{Application to Dark-Matter only Simulations} \label{sec:application}
We apply our SAM to 10 cosmological N-body simulations produced using {\sc gadget2} \citep{Springel2001}. Each simulation is 3 Mpc on a side and has $512^3$ particles corresponding to a particle mass of $8 \times 10^{3}~M_{\mathrm{\odot}}$. In each simulation, $\sim100$ snapshots were saved from $z=40$ to $z=6$, spaced in cosmic time by $\Delta t = t_{\mathrm{H}}/40$, where $t_{\mathrm{H}}$ is the Hubble time at the previous snapshot. This time spacing corresponds to $\approx 0.25$ of the dynamical time of a halo at its virial radius.

Within each simulation, we identify the number density of active Pop III sources, which we define as halos that have formed stars within the past 3 Myr. To ensure that we remain within the calibration regime of the Renaissance and \Aeos simulations, we impose a conservative halo mass ceiling of $10^{9}\, M_{\mathrm{\odot}}$. Once this threshold is reached, the SAM halts. Across the 10 realizations, our SAM evolves self-consistently down to $z\sim 10$, at which point the upper halo mass limit is reached.

\begin{figure}
\epsscale{1.1}
\plotone{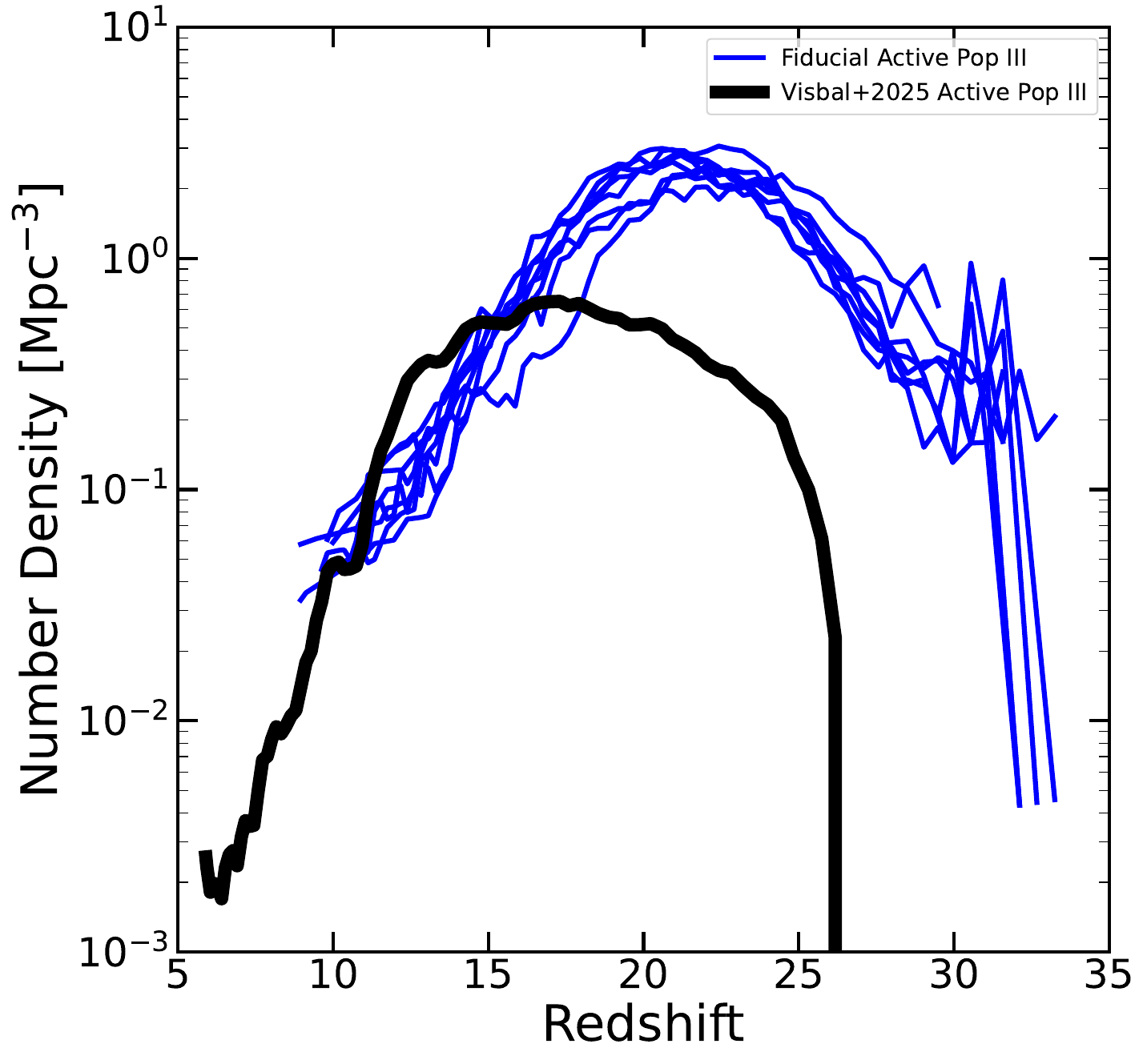}
\caption{Number density of Pop III sources younger than 3 Myr as a function of redshift for the fiducial SAM realizations run on the 10 N-body simulations is shown by the {\em blue lines}. The {\em black line} is the number density of Pop III sources younger than 3 Myr from \citet{Visbal2025b} that was compared to observations of LAP1-B using a prior implementation of the SAM. \label{fig:stellar_num_dens}}
\end{figure}

Figure \ref{fig:stellar_num_dens} shows the number densities of active Pop III sources with ages less than 3 Myr predicted by our fiducial SAM and the results of \citet{Visbal2025b} that compared a prior implementation of the SAM with observations of LAP1-B, a Pop III candidate system at $z = 6.6$ which is gravitationally lensed by the galaxy cluster MACS J041 \citep{Vanzella2023,Nakajima2025}. Both models were applied to the same 10 N-body simulations.

Our updated SAM predicts higher number densities of active Pop III sources than \citet{Visbal2025b} for $z > 15$. This difference arises because \citet{Visbal2025b} uses $M_{\mathrm{crit}}$ from \citet{Kulkarni2021} and incorporates a non-zero $v_{\mathrm{bc}}$. This results in delayed Pop III star formation at high redshift. At $z = 10$ our SAM produces a number density of active Pop III sources of $\sim 0.4~{\rm Mpc}^{-3}$. Following from  \citet{Visbal2025b} for LAP1-B, this corresponds to $\sim 10$ Pop III sources formed within 3 Myr at $z = 10$ that are within the lensed area where the foreground galaxy cluster MACS J0416 provides a magnification of $\mu > 30$.

\section{Conclusions}
\label{sec:conclusions}
In this work, we present a SAM for the formation of the first stars and galaxies, calibrated against two complementary cosmological hydrodynamical simulations. The \Aeos simulation resolves individual Pop III stars and their feedback in unprecedented detail, while the Renaissance simulations follow metal-enriched Pop II star formation in a variety of environments to $z \sim 11$. By calibrating to both, our SAM is uniquely positioned to capture the small-scale physics of primordial star formation and span a wide range in halo mass. This dual calibration allows us to build a predictive framework that bridges the gap between local star formation processes and the global history of the earliest galaxies.

A key result of our study is the close agreement between the SAM and \Aeos for Pop III star formation. Across 10 realizations of our fiducial model, the SAM accurately reproduces the evolution of total Pop III stellar mass and the number of halos hosting Pop III stars, with particularly strong agreement at $z>17$. Where small discrepancies arise, for example, the underestimation of Pop III stellar mass and halo counts at $z<17$, we identify their likely physical origin in local reionization effects. Importantly, the halos that form Pop III stars in \Aeos are predominantly the same halos that do so in the SAM, underscoring the physical fidelity of our approach. The large difference in Pop II stellar mass between the SAM and \Aeos for $z<17$ is likely due to the small number ($<20$) of \Aeos halos hosting Pop II star formation, which limits the robustness of our calibration to \Aeos. A couple of \Aeos halos having longer values of the time delay between the death of Pop III stars and the onset of Pop II formation in a given halo $t_{\mathrm{delay}}$ than the SAM could produce the difference. Despite the discrepancy in Pop II stellar mass, the number of Pop II-forming halos predicted by the fiducial models agrees well with \Aeos.

Another central finding is the necessity of modifying the critical mass threshold $M_{\mathrm{crit}}$ used to regulate Pop III star formation if a scatter is applied to this criterion similar to that seen in \Aeos. We demonstrate that without this modification, applying a scatter to $M_{\mathrm{crit}}$ in the SAM produces an artificial excess of low-mass halos forming Pop III stars at high redshift because there are more low-mass halos below the cutoff than higher mass halos above it. By applying our corrected prescription, we achieve significantly better agreement with \Aeos in both the total stellar mass and the number of star-forming halos. This highlights that the function for $M_{\mathrm{crit}}$ presented by \citet{Kulkarni2021} cannot be applied directly in SAMs without adjustment. Our modification offers a general solution for future applications.

We also address an important nuance in calibrating $t_{\mathrm{delay}}$. When relying on simulation outputs, it is crucial to account not only for halos where Pop II stars have already formed, but also for halos that have hosted Pop III stars and are still awaiting Pop II star formation, either because of intrinsically long $t_{\mathrm{delay}}$ values or because the simulation ends before the transition occurs. Our SAM incorporates this by implementing a simple prescription for $t_{\mathrm{delay}}$ that successfully captures the expected Pop III to Pop II transition and produces a delay time distribution consistent with the \Aeos sample. However, we find that most of the halos hosting Pop II stars in the SAM form in different halos compared to \Aeos, reflecting a shortcoming in our implementation of  $t_{\mathrm{delay}}$. Because the actual distribution of $t_{\mathrm{delay}}$ is not fully captured in \Aeos, our model cannot use a physically motivated calibration for this parameter. The importance of $t_{\mathrm{delay}}$ to the overall star formation history motivates dedicated simulations extending on the work of \citep{Jeon2014}.

Finally, we extend our SAM beyond the resolution and temporal limits of \Aeos by applying it to DM-only simulations. This allows us to make predictions for the abundance of halos hosting Pop III stars down to $z \sim 10$, reaching redshifts directly relevant for comparison with JWST observations of candidate Pop III galaxies. In agreement with \citet{Visbal2025b}, at $z=10$ we find that roughly ten Pop III sources formed within the past 3 Myr lie within the area lensed by the galaxy cluster MACS J0416 where the magnification exceeds $\mu>30$. This finding provides additional support for the prospect that both strongly lensed Pop III galaxies (e.g., LAP1-B) and individual Pop III stars may be detectable with JWST. However, our use of a streaming velocity $v_{\mathrm{bc}} = 0 \ \text{km s}^{-1}$ to match \Aeos is not representative of the conditions in a typical patch of the universe of similar size, which at recombination will have $v_{\mathrm{bc}} = 30 \ \text{km s}^{-1}$ \citep{Tseliakhovich2010}.

In summary, we describe a SAM that is (1) calibrated to two complementary hydrodynamical simulations, (2) physically consistent with small-scale primordial star formation, and (3) applicable to large-volume DM-only simulations. Unlike full hydrodynamical runs, our SAM can be evolved efficiently to lower redshifts directly relevant for JWST, allowing predictions to be compared against candidate Pop III galaxies. Its flexibility enables systematic tests of different prescriptions for star formation, reionization, and metal enrichment, and it can be applied to volumes large enough to begin probing the cosmological environment of primordial star formation. Critically, the SAM executes in a small fraction of the time required for full hydrodynamical simulations, running in less than a day on a laptop, while still incorporating the more realistic physics of the hydrodynamical simulations through calibration. This combination of speed, physical fidelity, and scalability makes our SAM a powerful framework for exploring the earliest stages of galaxy formation.

\section*{Acknowledgments}

R.H. and E.V. acknowledge support from NASA ATP grant 80NNSSC22K0629 and NSF grant AST-2009309. J.M. acknowledges support from the NSF Graduate Research Fellowship Program through grant DGE-2036197.  M.-M.M.L. acknowledges support from NSF grant AST23-07950. E.P.A. and M.-M.M.L. acknowledge support from NASA Astrophysical Theory Program grant 80NSSC24K0935. GLB acknowledges support from the NSF (AST-2108470, AST-2307419), NASA TCAN award 80NSSC21K1053, and the Simons Foundation through the Learning the Universe Collaboration. J.H.W. acknowledges support from NSF grants AST-2108020 and AST-2510197 and NASA grant 80NSSC21K1053.

The numerical simulations in this paper were run using the Ohio Supercomputer Center (OSC).  The authors also acknowledge the Texas Advanced Computing Center at The University of Texas at Austin for providing HPC and storage resources that have contributed to \Aeos.

\vspace{5mm}
\facilities{Ohio Supercomputer Center (OSC), Texas Advanced Comupting Center (TACC)}

\software{\textsc{numpy} \citep{NumPy2020}, 
          \textsc{scipy} \citep{SciPy2020},
          \textsc{matplotlib} \citep{Hunter2007},
          \textsc{cython} \citep{Cython2011}, 
          Jupyter notebook  \citep{Jupyter2016}, 
          \textsc{rockstar} \citep{2013behroozi_a}, 
          \textsc{consistent trees} \citep{2013behroozi_b}, 
          \textsc{yt} \citep{yt2011}, 
          \textsc{ytree} \citep{ytree2019}.
          }

\bibliographystyle{yahapj}
\bibliography{refs}

\begin{thebibliography}{}
\providecommand\natexlab[1]{#1}
\providecommand\JournalTitle[1]{#1}

\bibitem[{{Abel} {et~al.}(2002){Abel}, {Bryan}, \& {Norman}}]{Abel2002}
{Abel}, T., {Bryan}, G.~L., \& {Norman}, M.~L. 2002, \href{http://dx.doi.org/10.1126/science.295.5552.93}{\JournalTitle{Science}, 295, 93}

\bibitem[{{Adamo} {et~al.}(2024){Adamo}, {Bradley}, {Vanzella}, {Claeyssens}, {Welch}, {Diego}, {Mahler}, {Oguri}, {Sharon}, {Abdurro'uf}, {Hsiao}, {Xu}, {Messa}, {Lassen}, {Zackrisson}, {Brammer}, {Coe}, {Kokorev}, {Ricotti}, {Zitrin}, {Fujimoto}, {Inoue}, {Resseguier}, {Rigby}, {Jim{\'e}nez-Teja}, {Windhorst}, {Hashimoto}, \& {Tamura}}]{Adamo+2024}
{Adamo}, A., {Bradley}, L.~D., {Vanzella}, E., {et~al.} 2024, \href{http://dx.doi.org/10.1038/s41586-024-07703-7}{\JournalTitle{\nat}, 632, 513}

\bibitem[{{Ahn} {et~al.}(2009){Ahn}, {Shapiro}, {Iliev}, {Mellema}, \& {Pen}}]{Ahn2009}
{Ahn}, K., {Shapiro}, P.~R., {Iliev}, I.~T., {Mellema}, G., \& {Pen}, U.-L. 2009, \href{http://dx.doi.org/10.1088/0004-637X/695/2/1430}{\JournalTitle{\apj}, 695, 1430}

\bibitem[{{Andersson} {et~al.}(2023){Andersson}, {Agertz}, {Renaud}, \& {Teyssier}}]{Andersson2023}
{Andersson}, E.~P., {Agertz}, O., {Renaud}, F., \& {Teyssier}, R. 2023, \href{http://dx.doi.org/10.1093/mnras/stad692}{\JournalTitle{\mnras}, 521, 2196}

\bibitem[{{Andersson} {et~al.}(2025){Andersson}, {Rey}, {Pontzen}, {Cadiou}, {Agertz}, {Read}, \& {Martin}}]{Andersson2025}
{Andersson}, E.~P., {Rey}, M.~P., {Pontzen}, A., {et~al.} 2025, \href{http://dx.doi.org/10.3847/1538-4357/ad99d6}{\JournalTitle{\apj}, 978, 129}

\bibitem[{{Behling} {et~al.}(2025){Behling}, {Hazlett}, {Kulkarni}, \& {Visbal}}]{Behling2025}
{Behling}, T., {Hazlett}, R., {Kulkarni}, M., \& {Visbal}, E. 2025, \href{http://dx.doi.org/10.48550/arXiv.2508.04808}{\JournalTitle{arXiv e-prints}, arXiv:2508.04808}

\bibitem[{{Behnel} {et~al.}(2011){Behnel}, {Bradshaw}, {Citro}, {Dalcin}, {Seljebotn}, \& {Smith}}]{Cython2011}
{Behnel}, S., {Bradshaw}, R., {Citro}, C., {et~al.} 2011, \href{http://dx.doi.org/10.1109/MCSE.2010.118}{\JournalTitle{Computing in Science and Engineering}, 13, 31}

\bibitem[{{Behroozi} {et~al.}(2013{\natexlab{a}}){Behroozi}, {Wechsler}, \& {Wu}}]{2013behroozi_a}
{Behroozi}, P.~S., {Wechsler}, R.~H., \& {Wu}, H.-Y. 2013{\natexlab{a}}, \href{http://dx.doi.org/10.1088/0004-637X/762/2/109}{\JournalTitle{\apj}, 762, 109}

\bibitem[{{Behroozi} {et~al.}(2013{\natexlab{b}}){Behroozi}, {Wechsler}, {Wu}, {Busha}, {Klypin}, \& {Primack}}]{2013behroozi_b}
{Behroozi}, P.~S., {Wechsler}, R.~H., {Wu}, H.-Y., {et~al.} 2013{\natexlab{b}}, \href{http://dx.doi.org/10.1088/0004-637X/763/1/18}{\JournalTitle{\apj}, 763, 18}

\bibitem[{{Brauer} {et~al.}(2025{\natexlab{a}}){Brauer}, {Emerick}, {Mead}, {Ji}, {Wise}, {Bryan}, {Mac Low}, {C{\^o}t{\'e}}, {Andersson}, \& {Frebel}}]{AeosMethods}
{Brauer}, K., {Emerick}, A., {Mead}, J., {et~al.} 2025{\natexlab{a}}, \href{http://dx.doi.org/10.3847/1538-4357/ada4a1}{\JournalTitle{\apj}, 980, 41}

\bibitem[{{Brauer} {et~al.}(2025{\natexlab{b}}){Brauer}, {Mead}, {Wise}, {Bryan}, {Mac Low}, {Ji}, {Emerick}, {Andersson}, {Frebel}, \& {C{\^o}t{\'e}}}]{Brauer2025b}
{Brauer}, K., {Mead}, J., {Wise}, J.~H., {et~al.} 2025{\natexlab{b}}, \href{http://dx.doi.org/10.48550/arXiv.2502.20433}{\JournalTitle{arXiv e-prints}, arXiv:2502.20433}

\bibitem[{{Bressan} {et~al.}(2012){Bressan}, {Marigo}, {Girardi}, {Salasnich}, {Dal Cero}, {Rubele}, \& {Nanni}}]{Bressan2012}
{Bressan}, A., {Marigo}, P., {Girardi}, L., {et~al.} 2012, \href{http://dx.doi.org/10.1111/j.1365-2966.2012.21948.x}{\JournalTitle{\mnras}, 427, 127}

\bibitem[{{Bromm} {et~al.}(2002){Bromm}, {Coppi}, \& {Larson}}]{Bromm2002}
{Bromm}, V., {Coppi}, P.~S., \& {Larson}, R.~B. 2002, \href{http://dx.doi.org/10.1086/323947}{\JournalTitle{\apj}, 564, 23}

\bibitem[{{Bromm} \& {Loeb}(2003)}]{Bromm2003}
{Bromm}, V., \& {Loeb}, A. 2003, \href{http://dx.doi.org/10.1038/nature02071}{\JournalTitle{\nat}, 425, 812}

\bibitem[{{Bromm} \& {Loeb}(2004)}]{Bromm2004}
---. 2004, \href{http://dx.doi.org/10.1016/j.newast.2003.12.006}{\JournalTitle{\na}, 9, 353}

\bibitem[{{Brummel-Smith} {et~al.}(2019){Brummel-Smith}, {Bryan}, {Butsky}, {Corlies}, {Emerick}, {Forbes}, {Fujimoto}, {Goldbaum}, {Grete}, {Hummels}, {Kim}, {Koh}, {Li}, {Li}, {Li}, {OShea}, {Peeples}, {Regan}, {Salem}, {Schmidt}, {Simpson}, {Smith}, {Tumlinson}, {Turk}, {Wise}, {Abel}, {Bordner}, {Cen}, {Collins}, {Crosby}, {Edelmann}, {Hahn}, {Harkness}, {Harper-Clark}, {Kong}, {Kritsuk}, {Kuhlen}, {Larrue}, {Lee}, {Meece}, {Norman}, {Oishi}, {Paschos}, {Peruta}, {Razoumov}, {Reynolds}, {Silvia}, {Skillman}, {Skory}, {So}, {Tasker}, {Wagner}, {Wang}, {Xu}, \& {Zhao}}]{Enzo2019}
{Brummel-Smith}, C., {Bryan}, G., {Butsky}, I., {et~al.} 2019, \href{http://dx.doi.org/10.21105/joss.01636}{\JournalTitle{The Journal of Open Source Software}, 4, 1636}

\bibitem[{{Bryan} {et~al.}(2014{\natexlab{a}}){Bryan}, {Norman}, {O'Shea}, {Abel}, {Wise}, {Turk}, {Reynolds}, {Collins}, {Wang}, {Skillman}, {Smith}, {Harkness}, {Bordner}, {Kim}, {Kuhlen}, {Xu}, {Goldbaum}, {Hummels}, {Kritsuk}, {Tasker}, {Skory}, {Simpson}, {Hahn}, {Oishi}, {So}, {Zhao}, {Cen}, {Li}, \& {The Enzo Collaboration}}]{Enzo2014}
{Bryan}, G.~L., {Norman}, M.~L., {O'Shea}, B.~W., {et~al.} 2014{\natexlab{a}}, \href{http://dx.doi.org/10.1088/0067-0049/211/2/19}{\JournalTitle{\apjs}, 211, 19}

\bibitem[{{Bryan} {et~al.}(2014{\natexlab{b}}){Bryan}, {Norman}, {O'Shea}, {Abel}, {Wise}, {Turk}, {Reynolds}, {Collins}, {Wang}, {Skillman}, {Smith}, {Harkness}, {Bordner}, {Kim}, {Kuhlen}, {Xu}, {Goldbaum}, {Hummels}, {Kritsuk}, {Tasker}, {Skory}, {Simpson}, {Hahn}, {Oishi}, {So}, {Zhao}, {Cen}, {Li}, \& {Enzo Collaboration}}]{2014Bryan}
---. 2014{\natexlab{b}}, \href{http://dx.doi.org/10.1088/0067-0049/211/2/19}{\JournalTitle{\apjs}, 211, 19}

\bibitem[{{Cai} {et~al.}(2025){Cai}, {Li}, {Cai}, {Wu}, {Yu}, {Dickinson}, {Sun}, {Fan}, {Wang}, {Cullen}, {Bian}, {Lin}, \& {Zou}}]{Cai2025}
{Cai}, S., {Li}, M., {Cai}, Z., {et~al.} 2025, \href{http://dx.doi.org/10.48550/arXiv.2507.17820}{\JournalTitle{arXiv e-prints}, arXiv:2507.17820}

\bibitem[{{Clark} {et~al.}(2011){Clark}, {Glover}, {Smith}, {Greif}, {Klessen}, \& {Bromm}}]{Clark2011}
{Clark}, P.~C., {Glover}, S. C.~O., {Smith}, R.~J., {et~al.} 2011, \href{http://dx.doi.org/10.1126/science.1198027}{\JournalTitle{Science}, 331, 1040}

\bibitem[{{Corazza} {et~al.}(2022){Corazza}, {Miranda}, \& {Wuensche}}]{Corazza2022}
{Corazza}, L.~C., {Miranda}, O.~D., \& {Wuensche}, C.~A. 2022, \href{http://dx.doi.org/10.1051/0004-6361/202244334}{\JournalTitle{\aap}, 668, A191}

\bibitem[{{Crosby} {et~al.}(2013){Crosby}, {O'Shea}, {Smith}, {Turk}, \& {Hahn}}]{Crosby2013}
{Crosby}, B.~D., {O'Shea}, B.~W., {Smith}, B.~D., {Turk}, M.~J., \& {Hahn}, O. 2013, \href{http://dx.doi.org/10.1088/0004-637X/773/2/108}{\JournalTitle{\apj}, 773, 108}

\bibitem[{{de Bennassuti} {et~al.}(2017){de Bennassuti}, {Salvadori}, {Schneider}, {Valiante}, \& {Omukai}}]{deBennassuti2017}
{de Bennassuti}, M., {Salvadori}, S., {Schneider}, R., {Valiante}, R., \& {Omukai}, K. 2017, \href{http://dx.doi.org/10.1093/mnras/stw2687}{\JournalTitle{\mnras}, 465, 926}

\bibitem[{{Dekel} \& {Rees}(1987)}]{Dekel1987}
{Dekel}, A., \& {Rees}, M.~J. 1987, \href{http://dx.doi.org/10.1038/326455a0}{\JournalTitle{\nat}, 326, 455}

\bibitem[{{Dijkstra} {et~al.}(2004){Dijkstra}, {Haiman}, {Rees}, \& {Weinberg}}]{Dijkstra2004}
{Dijkstra}, M., {Haiman}, Z., {Rees}, M.~J., \& {Weinberg}, D.~H. 2004, \href{http://dx.doi.org/10.1086/380603}{\JournalTitle{\apj}, 601, 666}

\bibitem[{{Emerick} {et~al.}(2019){Emerick}, {Bryan}, \& {Mac Low}}]{Emerick2019}
{Emerick}, A., {Bryan}, G.~L., \& {Mac Low}, M.-M. 2019, \href{http://dx.doi.org/10.1093/mnras/sty2689}{\JournalTitle{\mnras}, 482, 1304}

\bibitem[{{Ezzeddine} {et~al.}(2019){Ezzeddine}, {Frebel}, {Roederer}, {Tominaga}, {Tumlinson}, {Ishigaki}, {Nomoto}, {Placco}, \& {Aoki}}]{Ezzeddine2019}
{Ezzeddine}, R., {Frebel}, A., {Roederer}, I.~U., {et~al.} 2019, \href{http://dx.doi.org/10.3847/1538-4357/ab14e7}{\JournalTitle{\apj}, 876, 97}

\bibitem[{{Feathers} {et~al.}(2024){Feathers}, {Kulkarni}, {Visbal}, \& {Hazlett}}]{Feathers2024}
{Feathers}, C.~R., {Kulkarni}, M., {Visbal}, E., \& {Hazlett}, R. 2024, \href{http://dx.doi.org/10.3847/1538-4357/ad1688}{\JournalTitle{\apj}, 962, 62}

\bibitem[{{Fernandez} {et~al.}(2014){Fernandez}, {Bryan}, {Haiman}, \& {Li}}]{Fernandez2014}
{Fernandez}, R., {Bryan}, G.~L., {Haiman}, Z., \& {Li}, M. 2014, \href{http://dx.doi.org/10.1093/mnras/stu230}{\JournalTitle{\mnras}, 439, 3798}

\bibitem[{{Frebel} \& {Norris}(2015)}]{FrebelNorris2015}
{Frebel}, A., \& {Norris}, J.~E. 2015, \href{http://dx.doi.org/10.1146/annurev-astro-082214-122423}{\JournalTitle{\araa}, 53, 631}

\bibitem[{{Fujimoto} {et~al.}(2025){Fujimoto}, {Naidu}, {Chisholm}, {Atek}, {Endsley}, {Kokorev}, {Furtak}, {Pan}, {Liu}, {Bromm}, {Venditti}, {Visbal}, {Sarmento}, {Weibel}, {Oesch}, {Brammer}, {Schaerer}, {Adamo}, {Berg}, {Bezanson}, {Bouwens}, {Chemerynska}, {Claeyssens}, {Dessauges-Zavadsky}, {Frebel}, {Korber}, {Labbe}, {Marques-Chaves}, {Matthee}, {McQuinn}, {Mu{\~n}oz}, {Natarajan}, {Saldana-Lopez}, {Suess}, {Volonteri}, \& {Zitrin}}]{Fujimoto2025}
{Fujimoto}, S., {Naidu}, R.~P., {Chisholm}, J., {et~al.} 2025, \href{http://dx.doi.org/10.3847/1538-4357/ade9a1}{\JournalTitle{\apj}, 989, 46}

\bibitem[{{Geha} {et~al.}(2013){Geha}, {Brown}, {Tumlinson}, {Kalirai}, {Simon}, {Kirby}, {VandenBerg}, {Mu{\~n}oz}, {Avila}, {Guhathakurta}, \& {Ferguson}}]{Geha2013}
{Geha}, M., {Brown}, T.~M., {Tumlinson}, J., {et~al.} 2013, \href{http://dx.doi.org/10.1088/0004-637X/771/1/29}{\JournalTitle{\apj}, 771, 29}

\bibitem[{{Gessey-Jones} {et~al.}(2022){Gessey-Jones}, {Sartorio}, {Fialkov}, {Mirouh}, {Magg}, {Izzard}, {de Lera Acedo}, {Handley}, \& {Barkana}}]{Gessey-Jones2022}
{Gessey-Jones}, T., {Sartorio}, N.~S., {Fialkov}, A., {et~al.} 2022, \href{http://dx.doi.org/10.1093/mnras/stac2049}{\JournalTitle{\mnras}, 516, 841}

\bibitem[{{Gnedin}(2000)}]{Gnedin2000}
{Gnedin}, N.~Y. 2000, \href{http://dx.doi.org/10.1086/317042}{\JournalTitle{\apj}, 542, 535}

\bibitem[{{Gnedin} \& {Hui}(1998)}]{Gnedin1998}
{Gnedin}, N.~Y., \& {Hui}, L. 1998, \href{http://dx.doi.org/10.1046/j.1365-8711.1998.01249.x}{\JournalTitle{\mnras}, 296, 44}

\bibitem[{{Greif} {et~al.}(2012){Greif}, {Bromm}, {Clark}, {Glover}, {Smith}, {Klessen}, {Yoshida}, \& {Springel}}]{Greif2012}
{Greif}, T.~H., {Bromm}, V., {Clark}, P.~C., {et~al.} 2012, \href{http://dx.doi.org/10.1111/j.1365-2966.2012.21212.x}{\JournalTitle{\mnras}, 424, 399}

\bibitem[{{Greif} {et~al.}(2011){Greif}, {Springel}, {White}, {Glover}, {Clark}, {Smith}, {Klessen}, \& {Bromm}}]{Greif2011}
{Greif}, T.~H., {Springel}, V., {White}, S. D.~M., {et~al.} 2011, \href{http://dx.doi.org/10.1088/0004-637X/737/2/75}{\JournalTitle{\apj}, 737, 75}

\bibitem[{{Gutcke} {et~al.}(2021){Gutcke}, {Pakmor}, {Naab}, \& {Springel}}]{Gutcke+2021}
{Gutcke}, T.~A., {Pakmor}, R., {Naab}, T., \& {Springel}, V. 2021, \href{http://dx.doi.org/10.1093/mnras/staa3875}{\JournalTitle{\mnras}, 501, 5597}

\bibitem[{{Haardt} \& {Madau}(2012)}]{HM2012}
{Haardt}, F., \& {Madau}, P. 2012, \href{http://dx.doi.org/10.1088/0004-637X/746/2/125}{\JournalTitle{\apj}, 746, 125}

\bibitem[{{Haiman} {et~al.}(2000){Haiman}, {Abel}, \& {Rees}}]{Haiman2000}
{Haiman}, Z., {Abel}, T., \& {Rees}, M.~J. 2000, \href{http://dx.doi.org/10.1086/308723}{\JournalTitle{\apj}, 534, 11}

\bibitem[{{Haiman} {et~al.}(1996){Haiman}, {Rees}, \& {Loeb}}]{Haiman1996-2}
{Haiman}, Z., {Rees}, M.~J., \& {Loeb}, A. 1996, \href{http://dx.doi.org/10.1086/177628}{\JournalTitle{\apj}, 467, 522}

\bibitem[{{Haiman} {et~al.}(1997){Haiman}, {Rees}, \& {Loeb}}]{Haiman1997}
---. 1997, \href{http://dx.doi.org/10.1086/303647}{\JournalTitle{\apj}, 476, 458}

\bibitem[{{Harris} {et~al.}(2020){Harris}, {Millman}, {van der Walt}, {Gommers}, {Virtanen}, {Cournapeau}, {Wieser}, {Taylor}, {Berg}, {Smith}, {Kern}, {Picus}, {Hoyer}, {van Kerkwijk}, {Brett}, {Haldane}, {del R{\'\i}o}, {Wiebe}, {Peterson}, {G{\'e}rard-Marchant}, {Sheppard}, {Reddy}, {Weckesser}, {Abbasi}, {Gohlke}, \& {Oliphant}}]{NumPy2020}
{Harris}, C.~R., {Millman}, K.~J., {van der Walt}, S.~J., {et~al.} 2020, \href{http://dx.doi.org/10.1038/s41586-020-2649-2}{\JournalTitle{\nat}, 585, 357}

\bibitem[{{Hartwig} {et~al.}(2018){Hartwig}, {Yoshida}, {Magg}, {Frebel}, {Glover}, {G{\'o}mez}, {Griffen}, {Ishigaki}, {Ji}, {Klessen}, {O'Shea}, \& {Tominaga}}]{Hartwig2018}
{Hartwig}, T., {Yoshida}, N., {Magg}, M., {et~al.} 2018, \href{http://dx.doi.org/10.1093/mnras/sty1176}{\JournalTitle{\mnras}, 478, 1795}

\bibitem[{{Hartwig} {et~al.}(2022){Hartwig}, {Magg}, {Chen}, {Tarumi}, {Bromm}, {Glover}, {Ji}, {Klessen}, {Latif}, {Volonteri}, \& {Yoshida}}]{ASLOTH}
{Hartwig}, T., {Magg}, M., {Chen}, L.-H., {et~al.} 2022, \href{http://dx.doi.org/10.3847/1538-4357/ac7150}{\JournalTitle{\apj}, 936, 45}

\bibitem[{{Hazlett} {et~al.}(2025){Hazlett}, {Kulkarni}, {Visbal}, \& {Wise}}]{Hazlett2025}
{Hazlett}, R., {Kulkarni}, M., {Visbal}, E., \& {Wise}, J.~H. 2025, \href{http://dx.doi.org/10.3847/1538-4357/ad919e}{\JournalTitle{\apj}, 978, 13}

\bibitem[{{Hegde} \& {Furlanetto}(2023)}]{HegdeFurlanetto2023}
{Hegde}, S., \& {Furlanetto}, S.~R. 2023, \href{http://dx.doi.org/10.1093/mnras/stad2308}{\JournalTitle{\mnras}, 525, 428}

\bibitem[{{Heger} \& {Woosley}(2010)}]{HegerWoosley2010}
{Heger}, A., \& {Woosley}, S.~E. 2010, \href{http://dx.doi.org/10.1088/0004-637X/724/1/341}{\JournalTitle{\apj}, 724, 341}

\bibitem[{{Hirano} {et~al.}(2015){Hirano}, {Hosokawa}, {Yoshida}, {Omukai}, \& {Yorke}}]{Hirano2015}
{Hirano}, S., {Hosokawa}, T., {Yoshida}, N., {Omukai}, K., \& {Yorke}, H.~W. 2015, \href{http://dx.doi.org/10.1093/mnras/stv044}{\JournalTitle{\mnras}, 448, 568}

\bibitem[{{Hirano} {et~al.}(2014){Hirano}, {Hosokawa}, {Yoshida}, {Umeda}, {Omukai}, {Chiaki}, \& {Yorke}}]{Hirano2014}
{Hirano}, S., {Hosokawa}, T., {Yoshida}, N., {et~al.} 2014, \href{http://dx.doi.org/10.1088/0004-637X/781/2/60}{\JournalTitle{\apj}, 781, 60}

\bibitem[{{Hoeft} {et~al.}(2006){Hoeft}, {Yepes}, {Gottl{\"o}ber}, \& {Springel}}]{Hoeft2006}
{Hoeft}, M., {Yepes}, G., {Gottl{\"o}ber}, S., \& {Springel}, V. 2006, \href{http://dx.doi.org/10.1111/j.1365-2966.2006.10678.x}{\JournalTitle{\mnras}, 371, 401}

\bibitem[{{Hosokawa} {et~al.}(2016){Hosokawa}, {Hirano}, {Kuiper}, {Yorke}, {Omukai}, \& {Yoshida}}]{Hosokawa2016}
{Hosokawa}, T., {Hirano}, S., {Kuiper}, R., {et~al.} 2016, \href{http://dx.doi.org/10.3847/0004-637X/824/2/119}{\JournalTitle{\apj}, 824, 119}

\bibitem[{{Hosokawa} {et~al.}(2011){Hosokawa}, {Omukai}, {Yoshida}, \& {Yorke}}]{Hosokawa2011}
{Hosokawa}, T., {Omukai}, K., {Yoshida}, N., \& {Yorke}, H.~W. 2011, \href{http://dx.doi.org/10.1126/science.1207433}{\JournalTitle{Science}, 334, 1250}

\bibitem[{{Hunter}(2007)}]{Hunter2007}
{Hunter}, J.~D. 2007, \href{http://dx.doi.org/10.1109/MCSE.2007.55}{\JournalTitle{Computing in Science and Engineering}, 9, 90}

\bibitem[{{Inayoshi} {et~al.}(2016){Inayoshi}, {Kashiyama}, {Visbal}, \& {Haiman}}]{Inayoshi2016}
{Inayoshi}, K., {Kashiyama}, K., {Visbal}, E., \& {Haiman}, Z. 2016, \href{http://dx.doi.org/10.1093/mnras/stw1431}{\JournalTitle{\mnras}, 461, 2722}

\bibitem[{{Ishiyama} \& {Hirano}(2025)}]{Ishiyama2025}
{Ishiyama}, T., \& {Hirano}, S. 2025, \href{http://dx.doi.org/10.48550/arXiv.2501.17540}{\JournalTitle{arXiv e-prints}, arXiv:2501.17540}

\bibitem[{{Jappsen} {et~al.}(2009){Jappsen}, {Klessen}, {Glover}, \& {Mac Low}}]{Jappsen2009}
{Jappsen}, A.-K., {Klessen}, R.~S., {Glover}, S. C.~O., \& {Mac Low}, M.-M. 2009, \href{http://dx.doi.org/10.1088/0004-637X/696/2/1065}{\JournalTitle{\apj}, 696, 1065}

\bibitem[{{Jaura} {et~al.}(2022){Jaura}, {Glover}, {Wollenberg}, {Klessen}, {Geen}, \& {Haemmerl{\'e}}}]{Jaura2022}
{Jaura}, O., {Glover}, S. C.~O., {Wollenberg}, K. M.~J., {et~al.} 2022, \href{http://dx.doi.org/10.1093/mnras/stac487}{\JournalTitle{\mnras}, 512, 116}

\bibitem[{{Jeon} \& {Ko}(2024)}]{JeonKo2024}
{Jeon}, M., \& {Ko}, M. 2024, \href{http://dx.doi.org/10.48550/arXiv.2411.17862}{\JournalTitle{arXiv e-prints}, arXiv:2411.17862}

\bibitem[{{Jeon} {et~al.}(2014){Jeon}, {Pawlik}, {Bromm}, \& {Milosavljevi{\'c}}}]{Jeon2014}
{Jeon}, M., {Pawlik}, A.~H., {Bromm}, V., \& {Milosavljevi{\'c}}, M. 2014, \href{http://dx.doi.org/10.1093/mnras/stu1980}{\JournalTitle{\mnras}, 444, 3288}

\bibitem[{{Ji} {et~al.}(2024){Ji}, {Curtis}, {Storm}, {Chandra}, {Schlaufman}, {Stassun}, {Heger}, {Pignatari}, {Price-Whelan}, {Bergemann}, {Stringfellow}, {Fr{\"o}hlich}, {Reggiani}, {Holmbeck}, {Tayar}, {Shah}, {Griffith}, {Laporte}, {Casey}, {Hawkins}, {Horta}, {Cerny}, {Thibodeaux}, {Usman}, {Amarante}, {Beaton}, {Cargile}, {Chiappini}, {Conroy}, {Johnson}, {Kollmeier}, {Li}, {Loebman}, {Meynet}, {Bizyaev}, {Brownstein}, {Gupta}, {Morrison}, {Pan}, {Ramirez}, {Rix}, \& {S{\'a}nchez-Gallego}}]{Ji2024}
{Ji}, A.~P., {Curtis}, S., {Storm}, N., {et~al.} 2024, \href{http://dx.doi.org/10.3847/2041-8213/ad19c4}{\JournalTitle{\apjl}, 961, L41}

\bibitem[{{Klessen} \& {Glover}(2023)}]{KlessenGlover2023}
{Klessen}, R.~S., \& {Glover}, S. C.~O. 2023, \href{http://dx.doi.org/10.1146/annurev-astro-071221-053453}{\JournalTitle{\araa}, 61, 65}

\bibitem[{{Kluyver} {et~al.}(2016){Kluyver}, {Ragan-Kelley}, {P{\'e}rez}, {Granger}, {Bussonnier}, {Frederic}, {Kelley}, {Hamrick}, {Grout}, {Corlay}, {Ivanov}, {Avila}, {Abdalla}, {Willing}, \& {Jupyter Development Team}}]{Jupyter2016}
{Kluyver}, T., {Ragan-Kelley}, B., {P{\'e}rez}, F., {et~al.} 2016, \href{http://dx.doi.org/10.3233/978-1-61499-649-1-87}{in IOS Press}, 87

\bibitem[{{Komiya} {et~al.}(2010){Komiya}, {Habe}, {Suda}, \& {Fujimoto}}]{Komiya2010}
{Komiya}, Y., {Habe}, A., {Suda}, T., \& {Fujimoto}, M.~Y. 2010, \href{http://dx.doi.org/10.1088/0004-637X/717/1/542}{\JournalTitle{\apj}, 717, 542}

\bibitem[{{Kroupa}(2001)}]{Kroupa2001}
{Kroupa}, P. 2001, \href{http://dx.doi.org/10.1046/j.1365-8711.2001.04022.x}{\JournalTitle{\mnras}, 322, 231}

\bibitem[{{Kulkarni} {et~al.}(2021){Kulkarni}, {Visbal}, \& {Bryan}}]{Kulkarni2021}
{Kulkarni}, M., {Visbal}, E., \& {Bryan}, G.~L. 2021, \href{http://dx.doi.org/10.3847/1538-4357/ac08a3}{\JournalTitle{\apj}, 917, 40}

\bibitem[{{Lah{\'e}n} {et~al.}(2020){Lah{\'e}n}, {Naab}, {Johansson}, {Elmegreen}, {Hu}, {Walch}, {Steinwand el}, \& {Moster}}]{Lahen2020}
{Lah{\'e}n}, N., {Naab}, T., {Johansson}, P.~H., {et~al.} 2020, \href{http://dx.doi.org/10.3847/1538-4357/ab7190}{\JournalTitle{\apj}, 891, 2}

\bibitem[{{Lah{\'e}n} {et~al.}(2023){Lah{\'e}n}, {Naab}, {Kauffmann}, {Sz{\'e}csi}, {Hislop}, {Rantala}, {Kozyreva}, {Walch}, \& {Hu}}]{Lahen+2023}
{Lah{\'e}n}, N., {Naab}, T., {Kauffmann}, G., {et~al.} 2023, \href{http://dx.doi.org/10.1093/mnras/stad1147}{\JournalTitle{\mnras}, 522, 3092}

\bibitem[{{Lake} {et~al.}(2024){Lake}, {Grudi{\'c}}, {Naoz}, {Yoshida}, {Williams}, {Burkhart}, {Marinacci}, {Vogelsberger}, \& {Chen}}]{Lake2024preprint}
{Lake}, W., {Grudi{\'c}}, M.~Y., {Naoz}, S., {et~al.} 2024, \href{http://dx.doi.org/10.48550/arXiv.2410.02868}{\JournalTitle{arXiv e-prints}, arXiv:2410.02868}

\bibitem[{{Lanz} \& {Hubeny}(2003)}]{Lanz2003}
{Lanz}, T., \& {Hubeny}, I. 2003, \href{http://dx.doi.org/10.1086/374373}{\JournalTitle{\apjs}, 146, 417}

\bibitem[{{Latif} {et~al.}(2013){Latif}, {Schleicher}, {Schmidt}, \& {Niemeyer}}]{Latif2013}
{Latif}, M.~A., {Schleicher}, D.~R.~G., {Schmidt}, W., \& {Niemeyer}, J. 2013, \href{http://dx.doi.org/10.1088/2041-8205/772/1/L3}{\JournalTitle{\apjl}, 772, L3}

\bibitem[{{Latif} {et~al.}(2022){Latif}, {Whalen}, \& {Khochfar}}]{Latif2022}
{Latif}, M.~A., {Whalen}, D., \& {Khochfar}, S. 2022, \href{http://dx.doi.org/10.3847/1538-4357/ac3916}{\JournalTitle{\apj}, 925, 28}

\bibitem[{{Liu} \& {Bromm}(2020)}]{LiuBromm2020}
{Liu}, B., \& {Bromm}, V. 2020, \href{http://dx.doi.org/10.1093/mnras/staa2143}{\JournalTitle{\mnras}, 497, 2839}

\bibitem[{{Liu} {et~al.}(2024){Liu}, {Hartwig}, {Sartorio}, {Dvorkin}, {Costa}, {Santoliquido}, {Fialkov}, {Klessen}, \& {Bromm}}]{Liu2024}
{Liu}, B., {Hartwig}, T., {Sartorio}, N.~S., {et~al.} 2024, \href{http://dx.doi.org/10.1093/mnras/stae2120}{\JournalTitle{\mnras}, 534, 1634}

\bibitem[{{Machacek} {et~al.}(2001){Machacek}, {Bryan}, \& {Abel}}]{Machacek2001}
{Machacek}, M.~E., {Bryan}, G.~L., \& {Abel}, T. 2001, \href{http://dx.doi.org/10.1086/319014}{\JournalTitle{\apj}, 548, 509}

\bibitem[{{Maiolino} {et~al.}(2024){Maiolino}, {{\"U}bler}, {Perna}, {Scholtz}, {D'Eugenio}, {Witten}, {Laporte}, {Witstok}, {Carniani}, {Tacchella}, {Baker}, {Arribas}, {Nakajima}, {Eisenstein}, {Bunker}, {Charlot}, {Cresci}, {Curti}, {Curtis-Lake}, {de Graaff}, {Egami}, {Ji}, {Johnson}, {Kumari}, {Looser}, {Maseda}, {Nelson}, {Robertson}, {Rodr{\'\i}guez Del Pino}, {Sandles}, {Simmonds}, {Smit}, {Sun}, {Venturi}, {Williams}, \& {Willmer}}]{Maiolino2024}
{Maiolino}, R., {{\"U}bler}, H., {Perna}, M., {et~al.} 2024, \href{http://dx.doi.org/10.1051/0004-6361/202347087}{\JournalTitle{\aap}, 687, A67}

\bibitem[{{Mead} {et~al.}(2025{\natexlab{a}}){Mead}, {Brauer}, {Bryan}, {Mac Low}, {Ji}, {Wise}, {Andersson}, {Frebel}, {Emerick}, \& {C{\^o}t{\'e}}}]{Mead+2025b}
{Mead}, J., {Brauer}, K., {Bryan}, G.~L., {et~al.} 2025{\natexlab{a}}, \href{http://dx.doi.org/10.48550/arXiv.2509.13580}{\JournalTitle{arXiv e-prints}, arXiv:2509.13580}

\bibitem[{{Mead} {et~al.}(2025{\natexlab{b}}){Mead}, {Brauer}, {Bryan}, {Mac Low}, {Ji}, {Wise}, {Emerick}, {Andersson}, {Frebel}, \& {C{\^o}t{\'e}}}]{Mead+2025a}
---. 2025{\natexlab{b}}, \href{http://dx.doi.org/10.3847/1538-4357/ada3c1}{\JournalTitle{\apj}, 980, 62}

\bibitem[{{Mirocha} {et~al.}(2018){Mirocha}, {Mebane}, {Furlanetto}, {Singal}, \& {Trinh}}]{Mirocha2018}
{Mirocha}, J., {Mebane}, R.~H., {Furlanetto}, S.~R., {Singal}, K., \& {Trinh}, D. 2018, \href{http://dx.doi.org/10.1093/mnras/sty1388}{\JournalTitle{\mnras}, 478, 5591}

\bibitem[{{Morishita} {et~al.}(2025){Morishita}, {Liu}, {Stiavelli}, {Treu}, {Bergamini}, \& {Zhang}}]{Morishita2025}
{Morishita}, T., {Liu}, Z., {Stiavelli}, M., {et~al.} 2025, \href{http://dx.doi.org/10.48550/arXiv.2507.10521}{\JournalTitle{arXiv e-prints}, arXiv:2507.10521}

\bibitem[{{Moriya} {et~al.}(2019){Moriya}, {Wong}, {Koyama}, {Tanaka}, {Oguri}, {Hilbert}, \& {Nomoto}}]{Moriya2019}
{Moriya}, T.~J., {Wong}, K.~C., {Koyama}, Y., {et~al.} 2019, \href{http://dx.doi.org/10.1093/pasj/psz035}{\JournalTitle{\pasj}, 71, 59}

\bibitem[{{Mowla} {et~al.}(2024){Mowla}, {Iyer}, {Asada}, {Desprez}, {Tan}, {Martis}, {Sarrouh}, {Strait}, {Abraham}, {Brada{\v{c}}}, {Brammer}, {Muzzin}, {Pacifici}, {Ravindranath}, {Sawicki}, {Willott}, {Estrada-Carpenter}, {Jahan}, {Noirot}, {Matharu}, {Rihtar{\v{s}}i{\v{c}}}, \& {Zabl}}]{Mowla+2024}
{Mowla}, L., {Iyer}, K., {Asada}, Y., {et~al.} 2024, \href{http://dx.doi.org/10.1038/s41586-024-08293-0}{\JournalTitle{\nat}, 636, 332}

\bibitem[{{Nakajima} {et~al.}(2025){Nakajima}, {Ouchi}, {Harikane}, {Vanzella}, {Ono}, {Isobe}, {Nishigaki}, {Tsujimoto}, {Nakamura}, {Xu}, {Umeda}, \& {Zhang}}]{Nakajima2025}
{Nakajima}, K., {Ouchi}, M., {Harikane}, Y., {et~al.} 2025, \href{http://dx.doi.org/10.48550/arXiv.2506.11846}{\JournalTitle{arXiv e-prints}, arXiv:2506.11846}

\bibitem[{{Nebrin} {et~al.}(2023){Nebrin}, {Giri}, \& {Mellema}}]{Nebrin2023}
{Nebrin}, O., {Giri}, S.~K., \& {Mellema}, G. 2023, \href{http://dx.doi.org/10.1093/mnras/stad1852}{\JournalTitle{\mnras}, 524, 2290}

\bibitem[{{Noh} \& {McQuinn}(2014)}]{Noh2014}
{Noh}, Y., \& {McQuinn}, M. 2014, \href{http://dx.doi.org/10.1093/mnras/stu1412}{\JournalTitle{\mnras}, 444, 503}

\bibitem[{{Okamoto} {et~al.}(2008){Okamoto}, {Gao}, \& {Theuns}}]{Okamoto2008}
{Okamoto}, T., {Gao}, L., \& {Theuns}, T. 2008, \href{http://dx.doi.org/10.1111/j.1365-2966.2008.13830.x}{\JournalTitle{\mnras}, 390, 920}

\bibitem[{{Omukai} {et~al.}(2005){Omukai}, {Tsuribe}, {Schneider}, \& {Ferrara}}]{Omukai2005}
{Omukai}, K., {Tsuribe}, T., {Schneider}, R., \& {Ferrara}, A. 2005, \href{http://dx.doi.org/10.1086/429955}{\JournalTitle{\apj}, 626, 627}

\bibitem[{{O'Shea} \& {Norman}(2008)}]{Oshea2008}
{O'Shea}, B.~W., \& {Norman}, M.~L. 2008, \href{http://dx.doi.org/10.1086/524006}{\JournalTitle{\apj}, 673, 14}

\bibitem[{{O'Shea} {et~al.}(2015){O'Shea}, {Wise}, {Xu}, \& {Norman}}]{2015oshea}
{O'Shea}, B.~W., {Wise}, J.~H., {Xu}, H., \& {Norman}, M.~L. 2015, \href{http://dx.doi.org/10.1088/2041-8205/807/1/L12}{\JournalTitle{\apjl}, 807, L12}

\bibitem[{{Parsons} {et~al.}(2022){Parsons}, {Mas-Ribas}, {Sun}, {Chang}, {Gonzalez}, \& {Mebane}}]{Parsons2022}
{Parsons}, J., {Mas-Ribas}, L., {Sun}, G., {et~al.} 2022, \href{http://dx.doi.org/10.3847/1538-4357/ac746b}{\JournalTitle{\apj}, 933, 141}

\bibitem[{{Planck Collaboration} {et~al.}(2014){Planck Collaboration}, {Ade}, {Aghanim}, {Armitage-Caplan}, {Arnaud}, {Ashdown}, {Atrio-Barandela}, {Aumont}, {Baccigalupi}, {Banday}, {Barreiro}, {Bartlett}, {Battaner}, {Benabed}, {Beno{\^\i}t}, {Benoit-L{\'e}vy}, {Bernard}, {Bersanelli}, {Bielewicz}, {Bobin}, {Bock}, {Bonaldi}, {Bond}, {Borrill}, {Bouchet}, {Bridges}, {Bucher}, {Burigana}, {Butler}, {Calabrese}, {Cappellini}, {Cardoso}, {Catalano}, {Challinor}, {Chamballu}, {Chary}, {Chen}, {Chiang}, {Chiang}, {Christensen}, {Church}, {Clements}, {Colombi}, {Colombo}, {Couchot}, {Coulais}, {Crill}, {Curto}, {Cuttaia}, {Danese}, {Davies}, {Davis}, {de Bernardis}, {de Rosa}, {de Zotti}, {Delabrouille}, {Delouis}, {D{\'e}sert}, {Dickinson}, {Diego}, {Dolag}, {Dole}, {Donzelli}, {Dor{\'e}}, {Douspis}, {Dunkley}, {Dupac}, {Efstathiou}, {Elsner}, {En{\ss}lin}, {Eriksen}, {Finelli}, {Forni}, {Frailis}, {Fraisse}, {Franceschi}, {Gaier}, {Galeotta}, {Galli}, {Ganga}, {Giard}, {Giardino}, {Giraud-H{\'e}raud},
  {Gjerl{\o}w}, {Gonz{\'a}lez-Nuevo}, {G{\'o}rski}, {Gratton}, {Gregorio}, {Gruppuso}, {Gudmundsson}, {Haissinski}, {Hamann}, {Hansen}, {Hanson}, {Harrison}, {Henrot-Versill{\'e}}, {Hern{\'a}ndez-Monteagudo}, {Herranz}, {Hildebrandt}, {Hivon}, {Hobson}, {Holmes}, {Hornstrup}, {Hou}, {Hovest}, {Huffenberger}, {Jaffe}, {Jaffe}, {Jewell}, {Jones}, {Juvela}, {Keih{\"a}nen}, {Keskitalo}, {Kisner}, {Kneissl}, {Knoche}, {Knox}, {Kunz}, {Kurki-Suonio}, {Lagache}, {L{\"a}hteenm{\"a}ki}, {Lamarre}, {Lasenby}, {Lattanzi}, {Laureijs}, {Lawrence}, {Leach}, {Leahy}, {Leonardi}, {Le{\'o}n-Tavares}, {Lesgourgues}, {Lewis}, {Liguori}, {Lilje}, {Linden-V{\o}rnle}, {L{\'o}pez-Caniego}, {Lubin}, {Mac{\'\i}as-P{\'e}rez}, {Maffei}, {Maino}, {Mandolesi}, {Maris}, {Marshall}, {Martin}, {Mart{\'\i}nez-Gonz{\'a}lez}, {Masi}, {Massardi}, {Matarrese}, {Matthai}, {Mazzotta}, {Meinhold}, {Melchiorri}, {Melin}, {Mendes}, {Menegoni}, {Mennella}, {Migliaccio}, {Millea}, {Mitra}, {Miville-Desch{\^e}nes}, {Moneti}, {Montier}, {Morgante},
  {Mortlock}, {Moss}, {Munshi}, {Murphy}, {Naselsky}, {Nati}, {Natoli}, {Netterfield}, {N{\o}rgaard-Nielsen}, {Noviello}, {Novikov}, {Novikov}, {O'Dwyer}, {Osborne}, {Oxborrow}, {Paci}, {Pagano}, {Pajot}, {Paladini}, {Paoletti}, {Partridge}, {Pasian}, {Patanchon}, {Pearson}, {Pearson}, {Peiris}, {Perdereau}, {Perotto}, {Perrotta}, {Pettorino}, {Piacentini}, {Piat}, {Pierpaoli}, {Pietrobon}, {Plaszczynski}, {Platania}, {Pointecouteau}, {Polenta}, {Ponthieu}, {Popa}, {Poutanen}, {Pratt}, {Pr{\'e}zeau}, {Prunet}, {Puget}, {Rachen}, {Reach}, {Rebolo}, {Reinecke}, {Remazeilles}, {Renault}, {Ricciardi}, {Riller}, {Ristorcelli}, {Rocha}, {Rosset}, {Roudier}, {Rowan-Robinson}, {Rubi{\~n}o-Mart{\'\i}n}, {Rusholme}, {Sandri}, {Santos}, {Savelainen}, {Savini}, {Scott}, {Seiffert}, {Shellard}, {Spencer}, {Starck}, {Stolyarov}, {Stompor}, {Sudiwala}, {Sunyaev}, {Sureau}, {Sutton}, {Suur-Uski}, {Sygnet}, {Tauber}, {Tavagnacco}, {Terenzi}, {Toffolatti}, {Tomasi}, {Tristram}, {Tucci}, {Tuovinen}, {T{\"u}rler}, {Umana},
  {Valenziano}, {Valiviita}, {Van Tent}, {Vielva}, {Villa}, {Vittorio}, {Wade}, {Wandelt}, {Wehus}, {White}, {White}, {Wilkinson}, {Yvon}, {Zacchei}, \& {Zonca}}]{Planck2014}
{Planck Collaboration}, {Ade}, P.~A.~R., {Aghanim}, N., {et~al.} 2014, \href{http://dx.doi.org/10.1051/0004-6361/201321591}{\JournalTitle{\aap}, 571, A16}

\bibitem[{{Qin} {et~al.}(2020){Qin}, {Mesinger}, {Park}, {Greig}, \& {Mu{\~n}oz}}]{Qin2020}
{Qin}, Y., {Mesinger}, A., {Park}, J., {Greig}, B., \& {Mu{\~n}oz}, J.~B. 2020, \href{http://dx.doi.org/10.1093/mnras/staa1131}{\JournalTitle{\mnras}, 495, 123}

\bibitem[{{Safarzadeh} \& {Haiman}(2020)}]{Safarzadeh2020}
{Safarzadeh}, M., \& {Haiman}, Z. 2020, \href{http://dx.doi.org/10.3847/2041-8213/abc253}{\JournalTitle{\apjl}, 903, L21}

\bibitem[{{Salpeter}(1955)}]{Salpeter1955}
{Salpeter}, E.~E. 1955, \href{http://dx.doi.org/10.1086/145971}{\JournalTitle{\apj}, 121, 161}

\bibitem[{{Schaerer}(2002)}]{Schaerer2002}
{Schaerer}, D. 2002, \href{http://dx.doi.org/10.1051/0004-6361:20011619}{\JournalTitle{\aap}, 382, 28}

\bibitem[{{Schauer} {et~al.}(2020){Schauer}, {Drory}, \& {Bromm}}]{Schauer2020}
{Schauer}, A. T.~P., {Drory}, N., \& {Bromm}, V. 2020, \href{http://dx.doi.org/10.3847/1538-4357/abbc0b}{\JournalTitle{\apj}, 904, 145}

\bibitem[{{Schauer} {et~al.}(2021){Schauer}, {Glover}, {Klessen}, \& {Clark}}]{Schauer2021}
{Schauer}, A. T.~P., {Glover}, S. C.~O., {Klessen}, R.~S., \& {Clark}, P. 2021, \href{http://dx.doi.org/10.1093/mnras/stab1953}{\JournalTitle{\mnras}, 507, 1775}

\bibitem[{{Shapiro} {et~al.}(1994){Shapiro}, {Giroux}, \& {Babul}}]{Shapiro1994}
{Shapiro}, P.~R., {Giroux}, M.~L., \& {Babul}, A. 1994, \href{http://dx.doi.org/10.1086/174120}{\JournalTitle{\apj}, 427, 25}

\bibitem[{{Sharda} {et~al.}(2020){Sharda}, {Federrath}, \& {Krumholz}}]{Sharda2020}
{Sharda}, P., {Federrath}, C., \& {Krumholz}, M.~R. 2020, \href{http://dx.doi.org/10.1093/mnras/staa1926}{\JournalTitle{\mnras}, 497, 336}

\bibitem[{{Sharda} \& {Menon}(2024)}]{ShardaMenon2024preprint}
{Sharda}, P., \& {Menon}, S.~H. 2024, \href{http://dx.doi.org/10.48550/arXiv.2405.18265}{\JournalTitle{arXiv e-prints}, arXiv:2405.18265}

\bibitem[{{Sk{\'u}lad{\'o}ttir} {et~al.}(2024){Sk{\'u}lad{\'o}ttir}, {Koutsouridou}, {Vanni}, {Amarsi}, {Lucchesi}, {Salvadori}, \& {Aguado}}]{Skuladottir2024}
{Sk{\'u}lad{\'o}ttir}, {\'A}., {Koutsouridou}, I., {Vanni}, I., {et~al.} 2024, \href{http://dx.doi.org/10.3847/2041-8213/ad4b1a}{\JournalTitle{\apjl}, 968, L23}

\bibitem[{{Sk{\'u}lad{\'o}ttir} {et~al.}(2021){Sk{\'u}lad{\'o}ttir}, {Salvadori}, {Amarsi}, {Tolstoy}, {Irwin}, {Hill}, {Jablonka}, {Battaglia}, {Starkenburg}, {Massari}, {Helmi}, \& {Posti}}]{Skuladottir2021}
{Sk{\'u}lad{\'o}ttir}, {\'A}., {Salvadori}, S., {Amarsi}, A.~M., {et~al.} 2021, \href{http://dx.doi.org/10.3847/2041-8213/ac0dc2}{\JournalTitle{\apjl}, 915, L30}

\bibitem[{{Smith} \& {Lang}(2019)}]{ytree2019}
{Smith}, B., \& {Lang}, M. 2019, \href{http://dx.doi.org/10.21105/joss.01881}{\JournalTitle{The Journal of Open Source Software}, 4, 1881}

\bibitem[{{Smith} \& {Sigurdsson}(2007)}]{Smith2007}
{Smith}, B.~D., \& {Sigurdsson}, S. 2007, \href{http://dx.doi.org/10.1086/518692}{\JournalTitle{\apjl}, 661, L5}

\bibitem[{{Smith} {et~al.}(2009){Smith}, {Turk}, {Sigurdsson}, {O'Shea}, \& {Norman}}]{Smith2009}
{Smith}, B.~D., {Turk}, M.~J., {Sigurdsson}, S., {O'Shea}, B.~W., \& {Norman}, M.~L. 2009, \href{http://dx.doi.org/10.1088/0004-637X/691/1/441}{\JournalTitle{\apj}, 691, 441}

\bibitem[{{Smith} {et~al.}(2017){Smith}, {Bryan}, {Glover}, {Goldbaum}, {Turk}, {Regan}, {Wise}, {Schive}, {Abel}, {Emerick}, {O'Shea}, {Anninos}, {Hummels}, \& {Khochfar}}]{GrackleMethod}
{Smith}, B.~D., {Bryan}, G.~L., {Glover}, S.~C.~O., {et~al.} 2017, \href{http://dx.doi.org/10.1093/mnras/stw3291}{\JournalTitle{\mnras}, 466, 2217}

\bibitem[{{Smith}(2021)}]{Smith2021}
{Smith}, M.~C. 2021, \href{http://dx.doi.org/10.1093/mnras/stab291}{\JournalTitle{\mnras}, 502, 5417}

\bibitem[{{Sobacchi} \& {Mesinger}(2013)}]{Sobacchi2013}
{Sobacchi}, E., \& {Mesinger}, A. 2013, \href{http://dx.doi.org/10.1093/mnrasl/slt035}{\JournalTitle{\mnras}, 432, L51}

\bibitem[{{Sodini} {et~al.}(2024){Sodini}, {D'Odorico}, {Salvadori}, {Vanni}, {Bischetti}, {Cupani}, {Davies}, {Becker}, {Ba{\~n}ados}, {Bosman}, {Davies}, {Paolo Farina}, {Ferrara}, {Keating}, {Kulkarni}, {Lai}, {Ryan-Weber}, {Maria Sebastian}, \& {Walter}}]{Sodini2024}
{Sodini}, A., {D'Odorico}, V., {Salvadori}, S., {et~al.} 2024, \href{http://dx.doi.org/10.1051/0004-6361/202349062}{\JournalTitle{\aap}, 687, A314}

\bibitem[{{Springel} {et~al.}(2001){Springel}, {Yoshida}, \& {White}}]{Springel2001}
{Springel}, V., {Yoshida}, N., \& {White}, S. D.~M. 2001, \href{http://dx.doi.org/10.1016/S1384-1076(01)00042-2}{\JournalTitle{\na}, 6, 79}

\bibitem[{{Stacy} {et~al.}(2010){Stacy}, {Greif}, \& {Bromm}}]{Stacy2010}
{Stacy}, A., {Greif}, T.~H., \& {Bromm}, V. 2010, \href{http://dx.doi.org/10.1111/j.1365-2966.2009.16113.x}{\JournalTitle{\mnras}, 403, 45}

\bibitem[{{Susa} {et~al.}(2014){Susa}, {Hasegawa}, \& {Tominaga}}]{Susa2014}
{Susa}, H., {Hasegawa}, K., \& {Tominaga}, N. 2014, \href{http://dx.doi.org/10.1088/0004-637X/792/1/32}{\JournalTitle{\apj}, 792, 32}

\bibitem[{{Tegmark} {et~al.}(1997){Tegmark}, {Silk}, {Rees}, {Blanchard}, {Abel}, \& {Palla}}]{Tegmark1997}
{Tegmark}, M., {Silk}, J., {Rees}, M.~J., {et~al.} 1997, \href{http://dx.doi.org/10.1086/303434}{\JournalTitle{\apj}, 474, 1}

\bibitem[{{Thoul} \& {Weinberg}(1996)}]{Thoul1996}
{Thoul}, A.~A., \& {Weinberg}, D.~H. 1996, \href{http://dx.doi.org/10.1086/177446}{\JournalTitle{\apj}, 465, 608}

\bibitem[{{Toma} {et~al.}(2016){Toma}, {Yoon}, \& {Bromm}}]{Toma2016}
{Toma}, K., {Yoon}, S.-C., \& {Bromm}, V. 2016, \href{http://dx.doi.org/10.1007/s11214-016-0250-7}{\JournalTitle{\ssr}, 202, 159}

\bibitem[{{Trenti} \& {Stiavelli}(2009)}]{Trenti2009}
{Trenti}, M., \& {Stiavelli}, M. 2009, \href{http://dx.doi.org/10.1088/0004-637X/694/2/879}{\JournalTitle{\apj}, 694, 879}

\bibitem[{{Tseliakhovich} \& {Hirata}(2010)}]{Tseliakhovich2010}
{Tseliakhovich}, D., \& {Hirata}, C. 2010, \href{http://dx.doi.org/10.1103/PhysRevD.82.083520}{\JournalTitle{\prd}, 82, 083520}

\bibitem[{{Turk} {et~al.}(2011){Turk}, {Smith}, {Oishi}, {Skory}, {Skillman}, {Abel}, \& {Norman}}]{yt2011}
{Turk}, M.~J., {Smith}, B.~D., {Oishi}, J.~S., {et~al.} 2011, \href{http://dx.doi.org/10.1088/0067-0049/192/1/9}{\JournalTitle{\apjs}, 192, 9}

\bibitem[{{Vanzella} {et~al.}(2023{\natexlab{a}}){Vanzella}, {Loiacono}, {Bergamini}, {Me{\v{s}}tri{\'c}}, {Castellano}, {Rosati}, {Meneghetti}, {Grillo}, {Calura}, {Mignoli}, {Brada{\v{c}}}, {Adamo}, {Rihtar{\v{s}}i{\v{c}}}, {Dickinson}, {Gronke}, {Zanella}, {Annibali}, {Willott}, {Messa}, {Sani}, {Acebron}, {Bolamperti}, {Comastri}, {Gilli}, {Caputi}, {Ricotti}, {Gruppioni}, {Ravindranath}, {Mercurio}, {Strait}, {Martis}, {Pascale}, {Caminha}, {Annunziatella}, \& {Nonino}}]{Vanzella2023}
{Vanzella}, E., {Loiacono}, F., {Bergamini}, P., {et~al.} 2023{\natexlab{a}}, \href{http://dx.doi.org/10.1051/0004-6361/202346981}{\JournalTitle{\aap}, 678, A173}

\bibitem[{{Vanzella} {et~al.}(2023{\natexlab{b}}){Vanzella}, {Claeyssens}, {Welch}, {Adamo}, {Coe}, {Diego}, {Mahler}, {Khullar}, {Kokorev}, {Oguri}, {Ravindranath}, {Furtak}, {Hsiao}, {Abdurro'uf}, {Mandelker}, {Brammer}, {Bradley}, {Brada{\v{c}}}, {Conselice}, {Dayal}, {Nonino}, {Andrade-Santos}, {Windhorst}, {Pirzkal}, {Sharon}, {de Mink}, {Fujimoto}, {Zitrin}, {Eldridge}, \& {Norman}}]{Vanzella+2023}
{Vanzella}, E., {Claeyssens}, A., {Welch}, B., {et~al.} 2023{\natexlab{b}}, \href{http://dx.doi.org/10.3847/1538-4357/acb59a}{\JournalTitle{\apj}, 945, 53}

\bibitem[{{Ventura} {et~al.}(2024){Ventura}, {Qin}, {Balu}, \& {Wyithe}}]{Ventura2024}
{Ventura}, E.~M., {Qin}, Y., {Balu}, S., \& {Wyithe}, J. S.~B. 2024, \href{http://dx.doi.org/10.1093/mnras/stae567}{\JournalTitle{\mnras}, 529, 628}

\bibitem[{{Virtanen} {et~al.}(2020){Virtanen}, {Gommers}, {Oliphant}, {Haberland}, {Reddy}, {Cournapeau}, {Burovski}, {Peterson}, {Weckesser}, {Bright}, {van der Walt}, {Brett}, {Wilson}, {Millman}, {Mayorov}, {Nelson}, {Jones}, {Kern}, {Larson}, {Carey}, {Polat}, {Feng}, {Moore}, {VanderPlas}, {Laxalde}, {Perktold}, {Cimrman}, {Henriksen}, {Quintero}, {Harris}, {Archibald}, {Ribeiro}, {Pedregosa}, {van Mulbregt}, \& {SciPy 1. 0 Contributors}}]{SciPy2020}
{Virtanen}, P., {Gommers}, R., {Oliphant}, T.~E., {et~al.} 2020, \href{http://dx.doi.org/10.1038/s41592-019-0686-2}{\JournalTitle{Nature Methods}, 17, 261}

\bibitem[{{Visbal} {et~al.}(2020){Visbal}, {Bryan}, \& {Haiman}}]{Visbal2020}
{Visbal}, E., {Bryan}, G.~L., \& {Haiman}, Z. 2020, \href{http://dx.doi.org/10.3847/1538-4357/ab994e}{\JournalTitle{\apj}, 897, 95}

\bibitem[{{Visbal} {et~al.}(2025{\natexlab{a}}){Visbal}, {Bryan}, \& {Haiman}}]{Visbal2025a}
---. 2025{\natexlab{a}}, \href{http://dx.doi.org/10.48550/arXiv.2506.14482}{\JournalTitle{arXiv e-prints}, arXiv:2506.14482}

\bibitem[{{Visbal} {et~al.}(2015{\natexlab{a}}){Visbal}, {Haiman}, \& {Bryan}}]{Visbal2015b}
{Visbal}, E., {Haiman}, Z., \& {Bryan}, G.~L. 2015{\natexlab{a}}, \href{http://dx.doi.org/10.1093/mnras/stv1941}{\JournalTitle{\mnras}, 453, 4456}

\bibitem[{{Visbal} {et~al.}(2015{\natexlab{b}}){Visbal}, {Haiman}, \& {Bryan}}]{Visbal2015a}
---. 2015{\natexlab{b}}, \href{http://dx.doi.org/10.1093/mnras/stv785}{\JournalTitle{\mnras}, 450, 2506}

\bibitem[{{Visbal} {et~al.}(2014){Visbal}, {Haiman}, {Terrazas}, {Bryan}, \& {Barkana}}]{Visbal2014}
{Visbal}, E., {Haiman}, Z., {Terrazas}, B., {Bryan}, G.~L., \& {Barkana}, R. 2014, \href{http://dx.doi.org/10.1093/mnras/stu1710}{\JournalTitle{\mnras}, 445, 107}

\bibitem[{{Visbal} {et~al.}(2025{\natexlab{b}}){Visbal}, {Hazlett}, \& {Bryan}}]{Visbal2025b}
{Visbal}, E., {Hazlett}, R., \& {Bryan}, G.~L. 2025{\natexlab{b}}, \href{http://dx.doi.org/10.48550/arXiv.2508.03842}{\JournalTitle{arXiv e-prints}, arXiv:2508.03842}

\bibitem[{{Wang} {et~al.}(2024){Wang}, {Cheng}, {Ge}, {Meng}, {Daddi}, {Yan}, {Ji}, {Jin}, {Jones}, {Malkan}, {Arrabal Haro}, {Brammer}, {Oguri}, {Hou}, \& {Zhang}}]{Wang2024}
{Wang}, X., {Cheng}, C., {Ge}, J., {et~al.} 2024, \href{http://dx.doi.org/10.3847/2041-8213/ad4ced}{\JournalTitle{\apjl}, 967, L42}

\bibitem[{{Windhorst} {et~al.}(2018){Windhorst}, {Timmes}, {Wyithe}, {Alpaslan}, {Andrews}, {Coe}, {Diego}, {Dijkstra}, {Driver}, {Kelly}, \& {Kim}}]{Windhorst2018}
{Windhorst}, R.~A., {Timmes}, F.~X., {Wyithe}, J. S.~B., {et~al.} 2018, \href{http://dx.doi.org/10.3847/1538-4365/aaa760}{\JournalTitle{\apjs}, 234, 41}

\bibitem[{{Wise} \& {Abel}(2007)}]{Wise2007}
{Wise}, J.~H., \& {Abel}, T. 2007, \href{http://dx.doi.org/10.1086/522876}{\JournalTitle{\apj}, 671, 1559}

\bibitem[{{Wise} {et~al.}(2012){Wise}, {Turk}, {Norman}, \& {Abel}}]{Wise2012a}
{Wise}, J.~H., {Turk}, M.~J., {Norman}, M.~L., \& {Abel}, T. 2012, \href{http://dx.doi.org/10.1088/0004-637X/745/1/50}{\JournalTitle{\apj}, 745, 50}

\bibitem[{{Xing} {et~al.}(2023){Xing}, {Zhao}, {Liu}, {Heger}, {Han}, {Aoki}, {Chen}, {Ishigaki}, {Li}, \& {Zhao}}]{Xing2023}
{Xing}, Q.-F., {Zhao}, G., {Liu}, Z.-W., {et~al.} 2023, \href{http://dx.doi.org/10.1038/s41586-023-06028-1}{\JournalTitle{\nat}, 618, 712}

\bibitem[{{Yong} {et~al.}(2021){Yong}, {Kobayashi}, {Da Costa}, {Bessell}, {Chiti}, {Frebel}, {Lind}, {Mackey}, {Nordlander}, {Asplund}, {Casey}, {Marino}, {Murphy}, \& {Schmidt}}]{Yong2021}
{Yong}, D., {Kobayashi}, C., {Da Costa}, G.~S., {et~al.} 2021, \href{http://dx.doi.org/10.1038/s41586-021-03611-2}{\JournalTitle{\nat}, 595, 223}

\bibitem[{{Yoshida} {et~al.}(2008){Yoshida}, {Omukai}, \& {Hernquist}}]{Yoshida2008}
{Yoshida}, N., {Omukai}, K., \& {Hernquist}, L. 2008, \href{http://dx.doi.org/10.1126/science.1160259}{\JournalTitle{Science}, 321, 669}

\end{thebibliography}

\end{document}